\documentclass[twocolumn]{autart}    

\usepackage{amsmath}
\usepackage{amssymb}
\usepackage{enumerate}
\pdfinclusioncopyfonts=1 \usepackage{graphicx}
\usepackage{xcolor}
\usepackage[round]{natbib}
\usepackage{times}
\usepackage{bm}
\usepackage{multirow}
\usepackage{diagbox}

\def\RR{{\mathbb R}}

\def\NN{\mathbb{N}}

\def\sign{\,\mathrm{sign}}

\def\QED{$\hfill\blacksquare$}
\def\mer{$\hfill\circ$}
\def\met{$\hfill\triangle$}

\newcommand{\abs}[1]{\left|#1\right|}

\newcommand{\changed}[1]{#1}

\newtheorem{lemma}{Lemma}[section]
\newtheorem{theorem}[lemma]{Theorem}
\newtheorem{defin}[lemma]{Definition}
\newtheorem{prp}[lemma]{Proposition}
\newtheorem{remark}[lemma]{Remark}

\begin{document}

\begin{frontmatter}

\title{Optimal Robust Exact Differentiation \\ via Linear Adaptive Techniques}

\thanks[footnoteinfo]{Work partially supported by the Christian Doppler Research Association, the Austrian Federal Ministry for Digital and Economic Affairs and the National Foundation for Research, Technology and
    Agencia I+D+i grant PICT 2018-01385, Argentina.}

\author[Austria]{Richard Seeber}\ead{richard.seeber@tugraz.at}, 
\author[Argentina]{Hernan Haimovich}\ead{haimovich@cifasis-conicet.gov.ar}

\address[Austria]{Graz University of Technology, Institute of Automation and Control, Christian Doppler Laboratory for Model Based Control of Complex Test Bed Systems, Graz, Austria. }

\address[Argentina]{Centro Internacional Franco-Argentino de 
  Ciencias de la Informaci\'on y de Sistemas (CIFASIS)
  CONICET-UNR, 2000 Rosario, Argentina}

\begin{keyword}
    Differentiation, Optimal worst-case accuracy, Discrete-time implementation
\end{keyword}

\begin{abstract}
The problem of differentiating a function with bounded second derivative in the presence of bounded measurement noise is considered in both continuous-time and sampled-data settings.
Fundamental performance limitations of causal differentiators, in terms of the smallest achievable worst-case differentiation error, are shown.
A robust exact differentiator is then constructed via the adaptation of a single parameter of a linear differentiator. 
It is demonstrated that the resulting differentiator exhibits a combination of properties that outperforms existing continuous-time differentiators: it is robust with respect to noise, it \emph{instantaneously} converges to the exact derivative in the absence of noise, and it attains the smallest possible---hence optimal---upper bound on its differentiation error under noisy measurements. 
For sample-based differentiators, the concept of quasi-exactness is introduced to classify differentiators that achieve the lowest possible worst-case error based on sampled measurements in the absence of noise.
A straightforward sample-based implementation of the proposed linear adaptive continuous-time differentiator is shown to achieve quasi-exactness after a single sampling step as well as a theoretically optimal differentiation error bound that, in addition, converges to the continuous-time optimal one as the sampling period becomes arbitrarily small.
A numerical simulation illustrates the presented formal results.
\end{abstract}

\end{frontmatter}

\section{Introduction}
\label{sec:intro}

The signal differentiation problem consists in estimating a signal's derivatives based on signal measurements. This seemingly simple task becomes complicated in the presence of measurement noise and other perturbations, even if measurements are assumed to be available continuously over time. In a practical setting, moreover, measurements become available only at sampling instants, and calculations need to be performed on a digital computer; this may present additional obstacles to differentiation.

Methods for signal differentiation include algebraic methods involving elementary differential algebraic (linear) operations \citep{mbojoi_na09}, linear high-gain observers \citep{vasiljevic2008error,khapra_ijrnc14} and sliding-mode differentiators \citep{yuxu_el96,levant1998robust,levant_ijc03}. The linear differentiators can have good measurement noise rejection capability but are not exact, meaning that in the absence of noise their output is not ensured to converge to the true value of the signal derivative. Sliding-mode differentiators, in contrast, can be exact but this may lead to the exact differentiation also of some (differentiable) noise signals. Some strategies even aim to combine the positive features of linear and sliding-mode differentiators \citep{ghabar_tac20}.

One specific differentiation problem of particular interest in relation to sliding-mode control is the estimation of a signal's first $n$ derivatives having knowledge of a bound for the $n+1$-th derivative.
In this context, the presence of measurement noise limits the accuracy of any differentiator with a bound depending on noise amplitude and known derivative bound. The expressions for these accuracy limitations are related to the so-called Landau-Kolmogorov inequalities \citep{kolmog_amst62,schcav_tr70}, cf. also \cite{levant2017sliding}. The differentiation accuracy for many sliding-mode differentiators can be shown to be of the asymptotic order indicated by these inequalities as the noise amplitude tends to zero \citep{levant_ijc03,cruzzavala2011uniform}.

Since the derivative estimates are usually required for achieving some subsequent control objective, an important feature of any differentiator is its convergence speed. Sliding-mode differentiators can converge in finite time \citep{levant1998robust}, whereas linear differentiators only do so asymptotically \citep{vasiljevic2008error}. Moreover, some sliding-mode differentiators can even converge in a finite time that is independent of initial conditions \citep{cruzzavala2011uniform}, a situation that is called fixed-time convergence \citep{bhaber_siamjco00}. In addition, such differentiators can be designed to achieve any given bound on the convergence time \citep{seehai_auto21}.

When only sampled measurements are available and differentiator implementation becomes digital, then the differentiators conceived as continuous-time systems must be discretized. Care must be taken, however, because with improper discretization the accuracy of the resulting discretized differentiator can worsen significantly compared to the continuous-time one \citep{levant_vss12}. The analysis of different discretization techniques becomes then highly relevant \citep{mojbro_ijrnc21,carsan_ijrnc21,andhor_icm21}. \citet{levant2014proper} propose a proper discretization that preserves the accuracy asymptotics of a continuous-time differentiator. Several extensions of this idea are given by \citet{barlev_auto20}. Moreover, \citet{hanlev_tac21} introduce a rather slight modification that allows to lower the resulting discrete differentiator's output chattering in the absence of noise. Filtering differentiators \citep{levyu_tac18,levliv_ejc20} are even capable of filtering out unbounded noise with small average values while preserving all advantageous features of the standard sliding-mode-based differentiators.
A radically different strategy for signal differentiation in a digital implementation setting is to directly consider the information carried by noisy sampled measurements accounting for noise magnitude and known derivative bound. If a suitable bound on the noise magnitude is known, then the differentiation problem can be performed through solving specific convex optimization problems \citep{haisee_lcss22} in the form of linear programs. This strategy is shown to achieve the best possible worst-case accuracy and to have explicitly computable fixed-time convergence and accuracy bounds for first-order differentiation, provided that the noise amplitude is known.

For the continuous-time case, an interesting consequence of the results in \citep{haisee_lcss22} is that the best possible worst-case accuracy may also be achieved by a linear finite difference.
However, this is true only if the noise magnitude is known to the differentiator; if the actual noise magnitude happens to be  lower than the assumed bound, then performance can worsen greatly.
In particular, such a linear differentiator can never be exact for arbitrary signals with bounded second derivative.
The use of such finite differences for robust output-feedback control was explored by \citet{levant2007finite}, who showed that when utilized in conjunction with homogeneous sliding-mode controllers, differentiation via finite differences turns into an appealing strategy by adaptation of the sampling time.
This adaptation allowed to achieve similar asymptotic accuracy as is obtained when using sliding-mode differentiators.
However, its use is restricted to differentiation in a closed feedback loop.

In this paper, first-order differentiation of noisy signals is considered.
Via adaptation of the parameter of a linear finite difference, a robust exact differentiator is constructed that achieves \emph{optimal differentiation accuracy}.
Contrary to what is called asymptotic optimality in \citep{levant1998robust}, which relates just to optimality of the exponents in the accuracy expression, optimality is achieved here in the sense that the differentiation error is ultimately bounded from above by the \emph{smallest possible upper bound}.
It is further shown that the convergence time of the resulting differentiator in the absence of noise is zero, i.e., that it outputs the true derivative from the beginning, and hence converges faster than any fixed-time differentiator.
For practical realization, it is shown that a straightforward sample-based realization achieves the lowest possible worst-case error among all sample-based differentiators in the absence of noise---a fact that is formalized by introducing the notion of quasi-exactness for sample-based differentiators.
Moreover, the proposed realization is shown to retain its optimal convergence behavior and its optimal accuracy with respect to the noise.

After a brief problem statement, Section~\ref{sec:problem} introduces formal definitions for worst-case differentiation error, exactness, and robustness that take the initial values of the signal into account.
Sections~\ref{sec:performance},~\ref{sec:lin-diff}, and~\ref{sec:adaptive-lin-diff} then consider the continuous-time case, while Sections~\ref{sec:sampled-performance} and~\ref{sec:sampled-adaptive-lin-diff} deal with the case of sampled measurements.
Sections~\ref{sec:performance} and~\ref{sec:sampled-performance}, in particular, show lower bounds for the worst-case error of all (causal) differentiators;
the former section discusses such performance limits for arbitrary and for exact differentiators, which motivate the notion of (optimal) differentiation accuracy, while the latter section shows performance limitations incurred by sampling, leading to the notion of quasi-exactness of sample-based differentiators.
As a preliminary result for the considered linear adaptive strategy, Section~\ref{sec:lin-diff} shows that a finite difference is an optimal causal linear differentiator, which is not exact, however.
The main results---the proposed optimal robust exact differentiators---are then presented in Sections~\ref{sec:adaptive-lin-diff} and~\ref{sec:sampled-adaptive-lin-diff} for the continuous-time and the sampled-data case, respectively; their formal properties, in particular, are summarized in the main Theorems~\ref{thm:proposed:exact} and~\ref{thm:proposed:quasi-exact}.
Section~\ref{sec:simulation}, finally, illustrates the results by means of a numerical simulation, and Section~\ref{sec:conclusion} draws conclusions.

\textbf{Notation:}
$\RR$, $\RR_{\ge 0}$, $\RR_{>0}$ denote the reals, nonnegative reals, and positive reals, respectively. $\NN$ denotes the natural numbers and $\NN_0$ the naturals including 0. 
If $\alpha\in\RR$, then $|\alpha|$ denotes its absolute value.
For a set $A$, the image of $A$ under $f$ is denoted by $f(A) = \{ f(a) : a \in A \}$.
One-sided limits of a function $f$ at time instant $T$ from above or below are written as $\lim_{t \to T^+} f(t)$ or $\lim_{t \to T^-} f(t)$, respectively. `Almost everywhere' is abbreviated as `a.e.'.

\section{Preliminaries and Problem Statement}
\label{sec:problem}

\newcommand{\F}{\mathcal{F}}
\newcommand{\FL}{\F_L}
\newcommand{\EN}{\mathcal{E}_N}
\newcommand{\meas}{u}
\newcommand{\fdothat}{\hat{\dot f}}
\newcommand{\Diff}{\mathcal{D}}
\newcommand{\Tc}{\mathcal{T}}
\newcommand{\CLg}{\overline C_L}
\newcommand{\CLa}{C_L}

This section introduces the considered problem of signal differentiation, and provides formal definitions for important properties of differentiators.

\subsection{Signal differentiation}

Consider the problem of computing the derivative of a function $f : \RR_{\ge 0} \to \RR$ based on a noisy measurement $\meas = f + \eta$, under the assumption that uniform bounds $N$ and $L$ for the noise $\eta$ and for the second derivative $\ddot f$, respectively, exist.
More precisely, let $\F$ denote the set of functions $f : \RR_{\ge 0} \to \RR$ such that $f$ is differentiable and $\dot f$ is Lipschitz continuous on $\RR_{\ge 0}$. Therefore, if $f\in\F$ then the second derivative $\ddot f$ exists almost everywhere (a.e.) due to Rademacher's Theorem. The corresponding classes of signals to consider, from which the measurements are generated, are hence given by
\begin{subequations}
    \begin{align}
        \FL = \{ f \in \F &: \abs{\ddot f(t)} \le L \text{ a.e. on } \RR_{\ge 0}\} \\
        \EN = \{ \eta \text{ measurable} &: \abs{\eta(t)} \le N \text{ for all }t\ge 0 \}.
    \end{align}
\end{subequations}
Write $\FL + \EN = \{ f + \eta : f \in \FL, \eta \in \EN\}$ for the set of inputs $u$ with fixed $L$ and $N$.
All possible inputs for the differentiator then belong to the set
\begin{equation}
    \mathcal U = \bigcup_{\substack{L \ge 0 \\ N \ge 0}} (\FL + \EN).
\end{equation}
Note that $\mathcal U$ includes all bounded measurable functions.

A differentiator is an operator $\Diff : \mathcal U \to ( \RR_{\ge 0} \to \RR)$ that maps the measured signal $\meas$ to an estimate $\Diff \meas$ for the derivative of $f$.
This estimate is hereafter denoted by $y = \Diff \meas$.

\begin{defin}
\label{def:causal:linear}
The differentiator $\Diff$ is said
\begin{itemize}
\item
    to be causal, if $[\Diff u_1](T) = [\Diff u_2](T)$ whenever $u_1(t) = u_2(t)$ for all $t \in [0, T]$\changed{;}
\item
    to be linear, if $\Diff (\alpha_1 u_1 + \alpha_2 u_2) = \alpha_1 \Diff u_1 + \alpha_2 \Diff u_2$ for $u_1, u_2 \in \mathcal U$ and $\alpha_1, \alpha_2 \in \RR$.
\end{itemize}
\end{defin}

Two further important properties are exactness and robustness, as introduced by \cite{levant1998robust}. 
To formally define these properties, the worst-case differentiation error is introduced next.

\subsection{Worst-case differentiation error}
\label{sec:definitions}

For future reference, for every $R\ge 0$, define the class of signals with bounded second derivative that, in addition, have a bounded initial value and initial derivative, as follows 
\begin{equation}
    \FL^R := \{ f \in \FL : |f(0)| \le R, |\dot f(0)| \le R \}.
\end{equation}
\begin{defin}
    \label{def:acc:abs}
    Let $L, N \in \RR_{\ge 0}$.
    A differentiator $\Diff$ is said to have
    \begin{itemize}
         \item
             worst-case error $M_N^f(t)$ from time $t\ge 0$ for a signal $f\in\FL$ with noise bound $N$ if
             \begin{equation}
                 M_N^f(t) = \sup_{\substack{u=f+\eta \\ \eta\in\EN}} \sup_{\tau \ge t} \big|\dot f(\tau) - [\Diff u](\tau)\big|\changed{;}
             \end{equation}
        \item
            worst-case error $M_N^{L,R}(t)$ from time $t\ge 0$ over the signal class $\FL^R$ with noise bound $N$ if
            \begin{equation}
                M_N^{L,R}(t) = \sup_{f\in\FL^R} M_N^f(t).
            \end{equation}
    \end{itemize}
\end{defin}
Clearly, $M^f_N(t)$ and hence also $M_N^{L,R}(t)$ are non-increasing with respect to $t$ and non-decreasing with respect to $N$.
The latter is also non-decreasing with respect to $L$ and $R$.

The worst-case error $M_N^{L,R}(t)$ for a fixed time instant $t$ may be an unbounded function of $R$.
This can be seen, e.g., by considering a linear differentiator and is the reason for restricting considerations to the signal class $\FL^R \subset \FL$ in Definition~\ref{def:acc:abs}.
Using the worst-case differentiation error from time $t$, different notions of exactness may now be defined.

\subsection{Exactness}

Exactness, as introduced by \cite{levant1998robust}, is an important property of differentiators.
It guarantees that noise-free signals in a certain class, specifically $\FL$ in the following, are differentiated exactly, i.e., that the worst-case differentiation error for the noise-free case, $M_0^{L,R}(t)$, is zero for certain values of $t$.
For a given differentiator, this property may sometimes be established only after a certain transient period.
Hence, distinctions between different convergence behaviors are made.

\begin{defin}
\label{def:exact}
    A differentiator $\Diff$ is said to be 
    \begin{itemize}
\item
            exact in finite time over $\FL$ if for every $R \ge 0$ there exists a $t_R$ such that $M_0^{L,R}(t_R) = 0$\changed{;}
        \item
            exact in fixed time over $\FL$ if there exists $t$ such that $M_0^{L,R}(t) = 0$ for all $R \ge 0$\changed{;}
        \item
            exact from the beginning over $\FL$ if $M_0^{L,R}(t) = 0$ for all $R \ge 0$ and all $t > 0$\changed{;}
        \item
            not exact over $\FL$ if it is not exact in finite time over $\FL$.
    \end{itemize}
\end{defin}

Note that the time instant $t = 0$ is excluded when speaking about exactness from the beginning, since a causal differentiator only knows $f(0)$ at this point.
This makes it impossible to deduce $\dot f(0)$; specifically, $M_0^{L,R}(0) \ge R$ for all causal differentiators.

For illustration purposes, consider the well-known robust exact differentiator (RED) proposed by \citet{levant1998robust}
\begin{align}
    \dot y_1 &= \changed{\lambda_1 L^{\frac{1}{2}}} |u-y_1|^{\frac{1}{2}} \sign(u - y_1) + y_2 & y_1(0) &= u(0) \nonumber \\
    \label{eq:red}
    \dot y_2 &= \changed{\lambda_2 L} \sign(u - y_1) & y_2(0) &= 0
\end{align}
with input $u$, output $y = y_2$, and constant positive parameters \changed{$\lambda_1$} and \changed{$\lambda_2 > 1$}.
With its solutions understood in the sense of \citet{filippov1988differential}, this differentiator is exact in finite time \changed{over $\FL$} for sufficiently large \changed{$\lambda_1$}.
In particular, for \changed{$\lambda_1 \ge \sqrt{8 \lambda_2}$, one has $t_R = \frac{R}{(\lambda_2 -1) L}$}, cf. \cite[Theorem 5]{seeber2018novel}.

While the RED is exact in finite time, the uniform robust exact differentiator (URED) proposed by \cite{cruzzavala2011uniform} is exact in fixed time.
In contrast, the linear high gain differentiator (HGD) in state-space form as considered by \cite{vasiljevic2008error} is not exact.
The only existing differentiator that is exact from the beginning is the Euler differentiator $\Diff u = \dot u$. It is well-defined only in the absence of noise, however, i.e., only for $u \in \FL$. In Section~\ref{sec:adaptive-lin-diff}, \changed{an original} differentiator is constructed that is exact from the beginning and is well-defined for input signals in $\mathcal{U}$.

\subsection{Robustness}

A concept that is closely connected with exactness is robustness.
In \cite{levant1998robust}, a differentiator is said to be robust if the differentiator output $\Diff (f + \eta)$ tends to $\Diff f$ uniformly as the uniform bound $N$ on the noise tends to zero.

To define robustness here, the following quantities (almost) analogous to $M_N^f(t)$ and $M_0^{L,R}(t)$ are introduced to quantify the deviation between $\Diff u$ and $\Diff f$:
\begin{align}
    Q^f_N(t) &= \sup_{\substack{u=f+\eta \\ \eta\in\EN}} \sup_{\tau \ge t} \big| [\Diff f](\tau) - [\Diff u](\tau)\big|, \\
    Q^{L,R}(t) &= \limsup_{N \to 0^+} \sup_{f \in \FL^R} Q^f_N(t).
\end{align}
With these abbreviations, robustness may then be defined in a similar style as exactness.

\begin{defin}
    A differentiator $\Diff$ is said to be
    \begin{itemize}
        \item robust in finite time over $\FL$, if for every $R \ge 0$ there exists a $t_R$ such that $Q^{L,R}(t_R) = 0$\changed{;}
        \item robust in fixed time over $\FL$, if there exists $t$ such that $Q^{L,R}(t) = 0$ for all $R  \ge 0$\changed{;}
    \item robust almost from the beginning over $\FL$, if
 $Q^{L,R}(t) = 0$ for all $R \ge 0$ and all $t > 0$\changed{;}
    \item robust from the beginning over $\FL$, if $Q^{L,R}(0) = 0$ for all $R \ge 0$\changed{;}
    \item
            not robust over $\FL$ if it is not robust in finite time over $\FL$.
    \end{itemize}
\end{defin}

Unlike for exactness, the time instant $t = 0$ is not excluded when speaking about robustness \emph{from the beginning}.
However, it is impossible for a causal differentiator to be both robust and exact from the beginning as shown in the following, which motivates the additional notion of robustness \emph{almost from the beginning}.
\changed{All differentiators previously mentioned, i.e., the HGD, the RED, and the URED, are robust from the beginning.}

\subsection{Exactness and robustness}

Some basic relationships between exactness and robustness are next established.
\begin{prp}
    Let $L \ge 0$ and let $\Diff$ be a causal differentiator.
    Then, $\Diff$ is either not robust from the beginning or not exact from the beginning over $\FL$.
\end{prp}
\begin{pf}
    Suppose to the contrary that $M_0^{L,1}(t) = 0$ for all $t > 0$ and that $Q^{L,0}(0) = 0$.
    Consider the function $g(t) = t$.
    Clearly, $g \in \FL^1$, and hence $[\Diff g](t) = 1$ for all $t > 0$ due to exactness.
    Let $N > 0$, and choose the signal $f = 0 \in\FL^0$ and  noise $\eta(t) = \min\{N,t\}$.
    This yields $u(t) = f(t) + \eta(t) = g(t)$ for $t \in [0,N]$, and hence $[\Diff u](N) = 1$.
    Consequently, $Q^f_N(0) \ge Q^f_N(N) \ge 1$, for all $N > 0$, which yields the contradiction $Q^{L,0}(0) \ge 1$.
    \QED
\end{pf}

\changed{The next proposition shows} the well-known fact \cite[cf.][]{levant1998robust} that linear differentiators cannot be both robust and exact over $\FL$ for $L > 0$.
\begin{prp}
\label{prop:linear:robustexact}
    Let $L > 0$ and let $\Diff$ be a linear differentiator.
    Then, $\Diff$ is either not robust or not exact over $\FL$.
\end{prp}
\begin{pf}
    Suppose to the contrary that $\Diff$ is robust and exact over $\FL$ in finite time.
    Let $T \ge 1$ be such that $M_0^{L,0}(T) = 0$ and $Q^{L,0}(T) = 0$.
    Let $\omega_k = 2 \pi k$ for $k = 1, 2, \ldots$ and consider the sequence of noise inputs $\eta_k(t) = \min\{1,t^2\} \sin (\omega_k t)/\omega_k$.
    One verifies that $\eta_k(0) = \dot \eta_k(0) = 0$.
    Since $L > 0$, there furthermore exists an $\alpha_k > 0$ for each $\eta_k$ such that $\alpha_k \eta_k \in \FL^{0}$, since $\ddot \eta_k$ is uniformly bounded.
    Due to linearity, $\Diff$ then differentiates also $\eta_k$ exactly for $t \ge T$.
    Thus, for $f = 0$,
    \begin{equation}
        Q^{f}_{1/\omega_k}(T) \ge  \sup_{\tau \ge T} |[\Diff \eta_k](\tau)| = \sup_{\tau \ge T} |\dot \eta_k(\tau)| = 1
    \end{equation}
    for all $k$.
    Since $1/\omega_k \to 0$ as $k \to \infty$, this contradicts $Q^{L,0}(T) = 0$.
    \QED
\end{pf}

\section{Performance Limits and Differentiation Accuracy}
\label{sec:performance}

To motivate the notion of optimal differentiation accuracy, worst-case error lower bounds for causal and causal exact differentiators are shown next.

\subsection{Worst-case error lower bound of causal differentiators}

The main limitation of the worst-case differentiation error of \emph{any} causal differentiator is shown in the following proposition.
It stems from the fact that the differentiator, when fed zero input $u = 0$, cannot distinguish between the zero function and a parabola arc $L t^2/2-N$ staying within the noise bound.
In Section~\ref{sec:lin-diff}, a differentiator is constructed that attains this lower bound.

\begin{prp}
    \label{prop:bound:causal}
    Let $\Diff$ be a causal differentiator and let $L,N,R,\tau \ge 0$.
    Then, $M_N^{L,R}(\tau) \ge 2\sqrt{NL}$.
\end{prp}

\begin{figure}
    \centering
    \includegraphics{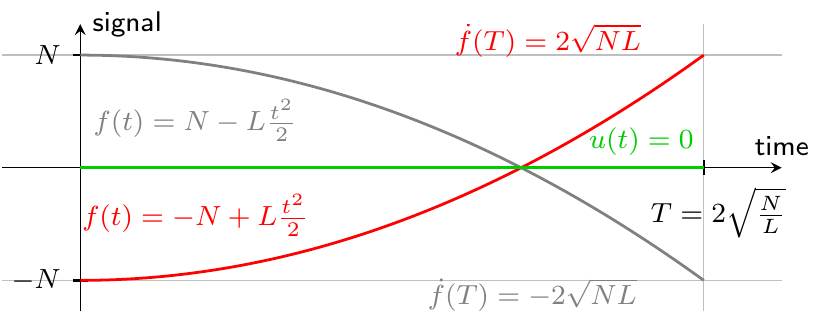}
    \caption{Input signal limiting the worst-case error of any causal differentiator: for zero input $u(t)$, no differentiator can distinguish between the two illustrated signals $f(t)$, leading to a worst-case error of at least $|\dot f(T)| = 2 \sqrt{N L}$ at time instant $T$.}
    \label{fig:causal_worst_case}
\end{figure}
\begin{remark}
\changed{The rationale of the proof is sketched in Fig.~\ref{fig:causal_worst_case}.}\mer
\end{remark}
\begin{pf}
    If $L = 0$ or $N = 0$, then there is nothing to prove.
    Otherwise, define $h_{\kappa} : [0, 2\kappa] \to [0, L\kappa^2]$ as
    \begin{equation}
    \label{eq:auxfunction:h}
        h(t) = \begin{cases}
            L \frac{t^2}{2} & t \in [0, \kappa) \\
            L \kappa^2 - L \frac{(t - 2 \kappa)^2}{2} & t \in [\kappa, 2\kappa]
        \end{cases}
    \end{equation}
    and let $\kappa = \sqrt{N/L}$.
    Then, $h_\kappa(0) = \dot h_\kappa(0) = \dot h_\kappa(2\kappa) = 0$ and $h_\kappa(2\kappa) = N$.
    Let $T = \tau + 4 \kappa$ and consider the functions
    \begin{align}
        g_1(t) &= \begin{cases}
            0 & t \in [0, \tau) \\
            -h_\kappa(t-\tau) & t \in [\tau, \tau+2\kappa) \\
            -N + L \frac{(t-\tau-2\kappa)^2}{2} & t \ge \tau + 2\kappa,
        \end{cases} \\
        g_2(t) &=
        \begin{cases}
            -g_1(t) & t \le T \\ -N & \text{otherwise}
        \end{cases}
    \end{align}
    It is straightforward to verify that for all $t\ge 0$, $\abs{\ddot g_1(t)} \le L$ a.e. and $\abs{g_2(t)} \le N$ hold, so that $g_1 \in \FL^0 \subseteq \FL^R$ and $g_2 \in \EN$.
    Choosing either $f = g_1$ and $\eta = g_2$ or $f = -g_1$ and $\eta = -g_2$ results in identical (zero) input for all $t \le T$.
    Since the differentiator is causal, its output value $y(T)$ must be the same in either case. 
    We have $\lim_{t\to T^-} 2 \dot g_1(t) = 4 \sqrt{N L}$. 
    Hence, $\max\{ M_N^{g_1}(\tau), M_N^{-g_1}(\tau) \} \ge 2\sqrt{N L}$ since $\tau \le T$ and, as a consequence, $M_N^{L,R}(\tau) \ge 2\sqrt{NL}$.
    \QED
\end{pf}

\subsection{Worst-case error lower bound of exact differentiators}

\begin{figure}
    \centering
    \includegraphics{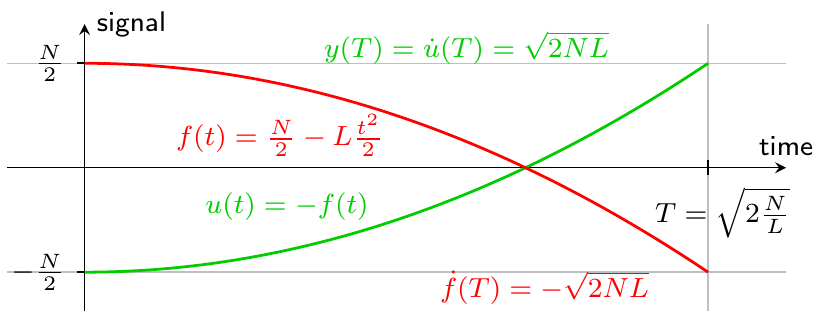}
    \caption{Input signal limiting the worst-case error of any causal and exact differentiator: for differentiable input $u(t) = -f(t)$, a causal exact differentiator makes an error of at least $2|\dot f(T)| = 2 \sqrt{2N L}$ at time instant $T$.}
    \label{fig:exact_worst_case}
\end{figure}

Exact differentiators have a more restrictive bound on their worst-case differentiation error than the bound in Proposition~\ref{prop:bound:causal}.
This is due to the fact that also some noise signals are differentiated exactly, as also noted by \citet{levant1998robust}.
Section~\ref{sec:adaptive-lin-diff} shows a differentiator that attains this bound.
\begin{prp}
\label{prop:bound:exact}
    Let $\Diff$ be a causal differentiator and let $L, N, R, \tau \ge 0$.
    If $\Diff$ is exact in finite time over $\FL$, then $M_N^{L,R}(\tau) \ge 2 \sqrt{2 N L}$.
\end{prp}
\begin{remark}
\changed{The rationale of the proof is sketched in Fig.~\ref{fig:exact_worst_case}.}\mer
\end{remark}
\begin{remark}
    A similar bound is shown in \cite[Proposition 1]{levant2017sliding} for arbitrary differentiation orders.
However, that bound is also valid for acausal differentiators, and hence the lower bound $2 \sqrt{N L}$ is obtained there for a first-order differentiator.
    \changed{A tighter} bound \changed{based on \cite{levant1998robust}} is mentioned by \cite{fraguela2012design} in a footnote.\mer
\end{remark}
\begin{pf}
    For $N = 0$ or $L = 0$, the statement is trivial.
Consider hence arbitrary $L, N > 0$, let $t_0$ be such that $M^{L,0}_0(t_0) = 0$ from Definition~\ref{def:exact}, and consider any $\tau \ge t_0$.
It will be shown that $M^{L,0}_N(\tau) \ge 2 \sqrt{2 N L}$, from which the claim follows from the fact that $M^{L,R}_N(t)$ is non-decreasing with respect to $R$ and non-increasing with respect to $t$.
Define $h_\kappa : [0, 2\kappa] \to [0, L\kappa^2]$ as in \eqref{eq:auxfunction:h} with $\kappa = \sqrt{N/L}$.
    Let $T = \tau + (2 + \sqrt{2}) \kappa$ and consider the functions
    \begin{align}
        \label{eq:g1-exact}
        g_1(t) &= \begin{cases}
            0 & t \in [0, \tau) \\
            - \frac{h_\kappa(t-\tau)}{2} & t \in [\tau, \tau + 2 \kappa) \\
            - \frac{N}{2} + L \frac{(t-\tau-2\kappa)^2}{2} & t \ge \tau + 2 \kappa,
        \end{cases} \\
        \label{eq:g2-exact}
        g_2(t) &=
        \begin{cases}
            2 g_1(t) & t \le T \\
            N & \text{otherwise}
        \end{cases}
    \end{align}
    It is straightforward to verify that $\abs{\ddot g_1(t)} \le L$ a.e. for $t\ge 0$, that $g_1 \in \FL^0$, and that $g_2 \in \EN$.
Due to exactness, $[\Diff g_1](t) = \dot g_1(t)$ for all $t \ge \tau$.
    Choosing $f = -g_1$ and $\eta = g_2$ yields $u(t) = f(t) + \eta(t) = g_1(t)$ for $t \in [0, T]$; hence, also $[\Diff u](t) = \dot g_1(t)$ for $t \in [\tau,T]$, since the differentiator is causal.
    Consequently, $[\Diff u](T) =  \sqrt{2 N L}$ whereas $\dot f(T) = -\sqrt{2 N L}$.
    This establishes that $M^f_N(\tau) \ge 2\sqrt{2NL}$ with $f\in\FL^0$. Therefore, $M^{L,0}_N(\tau) \ge 2\sqrt{2NL}$.
    \QED
\end{pf}

\subsection{Differentiation accuracy}

As one can see from Proposition~\ref{prop:bound:causal}, the lower bound of the worst-case differentiation error is proportional to $\sqrt{N L}$.
In \cite{vasiljevic2008error}, the HGD is also shown to achieve a similar proportionality with properly chosen parameters which, however, depend on $N$.

If the noise bound is unknown, it is desirable that this proportionality be maintained either for all or sufficiently small $N$.
This motivates the following notions of global or asymptotic  differentiation accuracy, respectively.
The latter notion is loosely consistent with similar notions in the context of sliding mode differentiatiors, where asymptotic accuracy also refers to the asymptotic behavior of the differentiation error with respect to (small) noise bounds $N$.

\begin{defin}
    \label{def:acc:asympt}
    \label{def:acc:global}
    A differentiator $\Diff$ is said to have
    \begin{enumerate}[a)]
    \item asymptotic accuracy $\CLa \in \RR_{\ge 0} \cup \{ \infty \}$ for signals in $\FL$, if $\CLa$ is the infimum of all numbers $C$ with the property
    that there exist $\epsilon > 0$ and a function $\Tc : \RR_{\ge 0} \times [0, \epsilon] \to \RR_{\ge 0}$ continuous in its second argument such that
    \begin{equation}
        \label{eq:rel-ac-bnd-single}
        M_N^{L,R}[\Tc(R,N)] \le C  \sqrt{N L}
    \end{equation}
    for all $N \in [0, \epsilon]$ and $R \ge 0$\changed{;}
    \item global accuracy $\CLg \in \RR_{\ge 0} \cup \{ \infty \}$ for signals in $\FL$, if $\CLg$ is the infimum of all numbers $C$ with the property
that there exists a function $\Tc : \RR_{\ge 0}^2 \to \RR_{\ge 0}$ continuous in its second argument such that \eqref{eq:rel-ac-bnd-single} holds for all $R, N \ge 0$.
    \end{enumerate}
\end{defin}

Note that $\Tc$ in this definition can be considered as a kind of convergence-time function.
For given initial value bound $R$ and noise bound $N$, it yields the time $\Tc(R, N)$ after which the worst-case differentiation error is bounded by $C \sqrt{N L}$. 
Asymptotic accuracy is a property pertaining to small values of the noise amplitude, whereas global accuracy applies to all noise amplitudes and is suitable when $N$ is unknown.
Clearly, $\CLg \ge \CLa$ holds by definition, and it is also possible that $\CLa$ is finite when $\CLg$ is infinite.
Important differentiators with a finite asymptotic accuracy are the RED and the URED.
The former also has finite global accuracy due to its homogeneity properties, while the latter does not.

The connection between robustness, exactness and accuracy is established by the following proposition, which shows that finite asymptotic accuracy implies robustness and exactness.

\begin{prp}
\label{prop:accuracy:exact}
    Let $L \ge 0$ and let $\Diff$ be a differentiator with asymptotic accuracy $\CLa \in \RR_{\ge 0}$.
    Then, $\Diff$ is both robust and exact in finite time over $\FL$.
\end{prp}\begin{remark}
From \eqref{eq:rel-ac-bnd-single}, one can see that $t_R = \Tc(R,0)$ in Definition~\ref{def:exact} with any $\Tc$ as in Definition~\ref{def:acc:asympt}. \mer
\end{remark}
\begin{pf}
    To prove robustness in finite time, note that $Q^f_N(t) \le M^f_N(t) + M^f_0(t)$ for all $\tau$ and $f \in \FL$.
    Let $R \ge 0$, $\epsilon > 0$, and let $\Tc$ be the function from Definition~\ref{def:acc:asympt} for some $C > \CLa$.
    Consider the image $\mathcal I = \Tc(R, [0, \epsilon])$, which is a compact interval due to continuity of $\Tc$ with respect to its second argument, and set $t_R = \max \mathcal I$.
    For all $N \in [0, \epsilon]$, then $M^{L,R}_N(t_R) \le C \sqrt{N L}$ since $M^{L,R}_N(\cdot)$ is non-increasing with respect to its argument.
    Hence,
    \begin{equation}
        \sup_{f \in \FL^R} Q^f_N(t_R) \le M^{L,R}_N(t_R) + M^{L,R}_0(t_R) \le C \sqrt{N L},
    \end{equation}
    holds for all $N \in [0, \epsilon]$, proving that $Q^{L,R}(t_R) = 0$.
    Exactness follows by noting that also $M^{L,R}_0(t_R) = 0$.
    \QED
\end{pf}

Note that, although only robustness in finite time may be concluded from finite asymptotic accuracy, most existing differentiators in fact exhibit robustness from the beginning.
In contrast, exactness from the beginning, as pointed out before, is a very strong property which is not known to be obtained by any existing robust differentiators up to now.

An immediate consequence of the previous proposition along with Proposition~\ref{prop:linear:robustexact} is the fact that linear differentiators cannot have finite asymptotic accuracy.
\begin{prp}
\label{prop:linear:accuracy}
    Let $L > 0$ and let $\Diff$ be a linear differentiator.
    Then, $\CLg = \CLa = \infty$.
\end{prp}
\begin{pf}
    Since $\CLg \ge \CLa$, assume to the contrary that $\CLa$ is finite.
    Then, $\Diff$ is robust and exact in finite time over $\FL$ by Proposition~\ref{prop:accuracy:exact}, contradicting Proposition~\ref{prop:linear:robustexact}.
    \QED
\end{pf}

From Proposition~\ref{prop:bound:causal} it is clear that $\CLa \ge 2$ bounds the asymptotic accuracy of any causal differentiator from below.
This bound is too conservative, however; a tighter lower bound, which will be shown to be achievable in Section~\ref{sec:adaptive-lin-diff}, is obtained from Propositions~\ref{prop:accuracy:exact} and~\ref{prop:bound:exact}.

\begin{prp}
    \label{prop:bound:accuracy}
Let $L > 0$ and let $\Diff$ be a causal differentiator.
    Then, $\CLg \ge \CLa \ge 2 \sqrt{2}$.
\end{prp}
\begin{pf}
It is sufficient to show $\CLa \ge 2 \sqrt{2}$.
To that end, assume to the contrary that $\CLa < 2 \sqrt{2}$.
    Then, $\CLa$ is finite and $\Diff$ is exact in finite time according to Proposition~\ref{prop:accuracy:exact}.
    From Proposition~\ref{prop:bound:exact}, then $M_N^{L,R}(t) \ge 2 \sqrt{2 N L}$ for all $N > 0$, yielding the contradiction $\CLa \ge 2 \sqrt{2}$.
    \QED
\end{pf}

For $L = 0$, finally, it is impossible to achieve finite accuracy.
\begin{prp}
\label{prop:accuracy:L0}
    Let $\Diff$ be a causal differentiator.
    Then, $\overline C_0 = C_0 = \infty$.
\end{prp}
\begin{pf}
    Suppose to the contrary that $C_L < \infty$ for $L = 0$ and let $R > 0$.
    From \eqref{eq:rel-ac-bnd-single}, then $N, \tau \in \RR_{>0}$ exist such that $M_N^{0,R}(\tau) = 0$.
    Let $\delta = \min\{N/\tau,R \}$ and consider $g(t) =  \delta t$.
    Clearly, $g \in \mathcal{F}_0^{R}$, and hence $[\Diff g](\tau) = \dot g(\tau) = \delta$.
    Now choose $f = 0 \in \mathcal{F}_0^{R}$ and $\eta(t) = \min\{ g(t), N\}$ to obtain $u(t) = f(t) + \eta(t) = g(t)$ for $t \le \tau$, leading to the contradiction $[\Diff u](\tau) - \dot f(\tau) = \delta$ due to causality, i.e., $M_N^{0,R}(\tau) \ge \delta > 0$.
    \QED
\end{pf}

\section{Linear differentiators with best worst-case error}
\label{sec:lin-diff}

As a preliminary result for ultimately constructing a robust exact differentiator with optimal differentiation accuracy, this section shows how a linear differentiator with lowest possible worst-case error $2\sqrt{NL}$, as given by Proposition~\ref{prop:bound:causal}, can be constructed.

Consider the following differentiator $\Diff$ with a positive parameter $T$ and output $y = \Diff u$ given by
\begin{equation}
    \label{eq:diff:linear}
    y(t) = \begin{cases}
        \frac{u(t) - u(t-T)}{T} & \text{if $t \ge T$} \\
        0 & \text{otherwise}.
    \end{cases}
\end{equation}
If the noise bound $N$ is precisely known, then this differentiator achieves the best possible worst-case differentiation error by suitable selection of $T$. 
\begin{theorem}
    \label{thm:knowN-lin-best}
    Let $N > 0$, $L > 0$ and consider the differentiator \eqref{eq:diff:linear} with $T = 2 \sqrt{N/L}$.
    Then, $M^{L,R}_N(\tau) = 2 \sqrt{N L}$ for all $R \ge 0$ and all $\tau \ge T$.\met
\end{theorem}

The proof uses the following two \changed{lemmata}.
\begin{lemma}
    \label{lem:f:changerate}
    Let $L \in \RR_{\ge 0}$ and $f \in \FL$.
    Then,
    \begin{equation}
        \label{eq:f:der:growth}
        \abs{f(t- \sigma) - f(t) + \dot f(t) \sigma} \le \frac{L \sigma^2}{2}
    \end{equation}
    holds for all $t \ge 0$ and all $\sigma \in [0, t]$.
\end{lemma}
\begin{pf}
    \changed{Consider an arbitrary $t \ge 0$ and} define the function
    $
        g(\sigma) = f(t- \sigma) - f(t) + \dot f(t) \sigma.
    $
    Clearly, $g$ is a.e. twice differentiable, its second derivative satisfies $\ddot g(\sigma) = \ddot f(t - \sigma)$ a.e. in $[0,t]$, and $g(0) = \dot g(0) = 0$.
    The lemma's claim $|g(\sigma)| \le L\sigma^2/2$ follows by double integration of $\ddot g$ using the bound $\abs{\ddot g(\sigma)} \le L$ starting from $\sigma = 0$.
    \QED
\end{pf}

\begin{lemma}
    \label{lem:diff:linear}
    Let $T > 0$ and consider the differentiator $\Diff$ with output $y = \Diff u$ defined in \eqref{eq:diff:linear}.
    Suppose that $f \in \FL$ and $\eta \in \EN$.
    Then,
    $
        |y(t) - \dot f(t)| \le \frac{2 N}{T} + \frac{L T}{2}
    $
    holds for all $t \ge T$.
\end{lemma}
\begin{remark}
As will be shown later, the bound is also valid when the parameter $T$ is time varying.
This motivates the adaptation of that parameter to eventually obtain a robust exact differentiator in Section~\ref{sec:adaptive-lin-diff}.
A related approach is proposed by \cite{levant2007finite}, where finite differences with adaptation of the sampling time are used in a closed-loop sliding mode control scheme.\mer
\end{remark}
\begin{pf}
    From \eqref{eq:diff:linear},
$\abs{u(t-T) - u(t) + y(t) T} = 0$
for $t \ge T$, and hence
    $
        \abs{f(t-T) - f(t) + y(t) T} \le 2N
    $, \changed{since $|u(\tau)-f(\tau)|\le N$ for all $\tau$}.
    Moreover, setting $\sigma = T$ in \eqref{eq:f:der:growth} in Lemma~\ref{lem:f:changerate} and combining the two inequalities yields
    $
        \abs{y(t) T - \dot f(t) T} \le L T^2/2 + 2 N.
    $
    The claim follows after dividing by $T$.
    \QED
\end{pf}

Using Lemma~\ref{lem:diff:linear}, the optimal worst-case accuracy of the linear differentiator \eqref{eq:diff:linear} may now be shown.

\noindent\textbf{PROOF of Theorem~\ref{thm:knowN-lin-best}.}
    From Lemma~\ref{lem:diff:linear} with parameter $T = 2 \sqrt{N/L}$, obtain
    \begin{equation}
        M_N^{L,R}(t) \le \frac{2N}{T} + \frac{L T}{2} = 2 \sqrt{N L}
    \end{equation}
    for $t \ge T$.
    Equality is concluded by noting that also $M_N^{L,R}(t) \ge 2 \sqrt{N L}$ due to Proposition~\ref{prop:bound:causal}.
    \QED

\section{Robust Exact Differentiators with Optimal Accuracy}
\label{sec:adaptive-lin-diff}

The problem with differentiator \eqref{eq:diff:linear} is that its tuning requires knowledge of the noise amplitude.
If the noise affecting its input is actually of lower amplitude than the design parameter $N$ used in the differentiator's construction, achieving an optimal worst-case differentiation error cannot be ensured.
The main idea for obtaining an optimal differentiator based on knowledge of only an upper bound on the noise amplitude or directly without any knowledge on the noise amplitude is to obtain a reasonable estimate $\hat N$ for this amplitude.

\newcommand{\gammamax}{\overline\gamma}
\newcommand{\Nmax}{\overline N}
\newcommand{\kmax}{\overline k}
\newcommand{\Tmax}{\overline T}
\subsection{Proposed differentiator}

From the above considerations, a robust exact differentiator with best possible worst-case accuracy is constructed as the linear time-varying differentiator\footnote{\changed{See Section~\ref{sec:sampled-adaptive-lin-diff} for the practical, sample-based implementation of the proposed continuous-time differentiator.}}
\begin{subequations}
    \label{eq:proposed:diff}
    \begin{align}
    \label{eq:proposed:diff:y}
        &y(t) = \begin{cases}
            0 & \text{if } t = 0 \\
            \lim_{T \to 0^+} \dfrac{u(t) - u(t - T)}{T} & \text{if } t > 0, \hat T(t) = 0 \\
            \dfrac{u(t) - u(t - \hat T(t))}{\hat T(t)} & \text{if } t > 0, \hat T(t) > 0.
        \end{cases} \\
        \intertext{with an adaptation of the time difference $\hat T(t)$ according to}
                \label{eq:def:That:t}
        &\hat T(t) =  \min\biggl\{t, \Tmax, 2 \gamma(t) \sqrt{ \frac{\hat N(t)}{L}}\biggr\}
        \displaybreak[0]
        \intertext{wherein the parameter $\Tmax \in \RR_{>0} \cup \{ \infty \}$ constrains the time difference from above, $\gamma :  \RR_{\ge 0} \to [1,\gammamax]$ is an arbitrary function constrained by a constant parameter $\gammamax \ge 1$, i.e.,}
        &\gamma(t) \in [1, \gammamax] \qquad \text{for all $t$,}
        \intertext{and $\hat N(t)$ is an estimate for the noise amplitude determined from the measurement $u$ according to}
        \label{eq:def:Nhat:t}
&\hat N(t) = \frac{1}{2} \sup_{\substack{T \in (0, \Tmax] \\ T \le t \\ \sigma \in [0, T]}} \biggl( \abs{Q(t, T, \sigma)} - \frac{L \sigma (T - \sigma)}{2} \biggr)\\
        \intertext{with the abbreviation $Q(t, T, \sigma)$ defined as}
        \label{eq:def:aT:t}
        &Q(t, T, \sigma) = u(t - \sigma) - u(t) + \frac{u(t) - u(t-T)}{T} \sigma. 
\end{align}
\end{subequations}
Note that $\hat N(t) \ge 0$ since $Q(t, T, 0) = 0$ and hence the argument of the supremum is zero for $\sigma = 0$.

The parameter $\Tmax$ may be considered to be a window-length parameter, since computing $y(t)$ requires evaluation of $u$ only on the interval $[t-\Tmax,t]$.
If a finite value is chosen for this parameter, it allows to limit the proposed differentiator's computational complexity.

The function $\gamma$ satisfying $1 \le \gamma(t) \le \gammamax$ for all $t$ is a degree of freedom parametrizing a whole family of optimal robust exact differentiators.
In the presence of sampled measurements, in Section~\ref{sec:sample-based-diff}, this degree of freedom will be exploited to choose $\hat T(t)$ as a multiple of the sampling time, yielding a straightforward discrete-time implementation.

The next main theorem, which is proven in Section~\ref{sec:proofs:continuous}, establishes that
(i) the output $y(t)$ of this differentiator is always well-defined and, in particular, the one-sided limit occuring in \eqref{eq:proposed:diff:y} exists whenever $\hat T(t) = 0$ and $t > 0$, (ii) this differentiator is exact from the beginning and robust almost from the beginning with any $\gammamax \ge 1$, and (iii) it has optimal \changed{asymptotic} accuracy $\CLa = 2 \sqrt{2}$ for $\gammamax \in [1, 1+\sqrt{2}]$ and also optimal global accuracy if moreover $\Tmax = \infty$.

\begin{theorem}
\label{thm:proposed:exact}
\label{thm:proposed:errorbound}
Let $L \in \RR_{>0}$ and consider the differentiator $\Diff$ with output $y = \Diff u$ defined by \eqref{eq:proposed:diff} with parameters $\gammamax \ge 1$ and $\Tmax \in \RR_{> 0} \cup \{ \infty \}$. Then, \changed{the following statements are true:}
\begin{enumerate}[a)]
    \item The output $y = \Diff u$ is well-defined for all $u \in \mathcal{U}$.\label{item:welldef}
    \item $\Diff$ is robust almost from the beginning and exact from the beginning over $\FL$.\label{item:exact}
    \item If $\gammamax \in [1, 1+\sqrt{2}]$, then $\Diff$ achieves optimal asymptotic accuracy $\CLa = 2 \sqrt{2}$; specifically, $M_N^{L,R}(t) \le 2 \sqrt{2 N L}$ holds for all $N \in [0, L \Tmax^2/2)$ and all $t > \sqrt{2 N/L}$.\label{item:errorbound}
    \item If $\gammamax \in [1, 1+\sqrt{2}]$ and $\Tmax = \infty$, then $\Diff$ achieves optimal global accuracy $\CLg = 2 \sqrt{2}$.\label{item:global}
    \met
    \end{enumerate}
\end{theorem}
\changed{\begin{remark}[Tuning]
\label{rem:tuning}
If a (crude) upper bound $\Nmax$ for the noise ampltiude is known, then a robust exact differentiator with bounded complexity that is optimal for all $N < \Nmax$ may be obtained by choosing the parameters $\gammamax = 1$ and $\Tmax = \sqrt{2 \Nmax/L}$.\mer
\end{remark}}\begin{remark}
The proposed differentiator achieves the best possible exactness and robustness features, 
given that it is impossible to achieve both properties from the beginning, according to Proposition~\ref{prop:linear:robustexact}, as well as the best possible accuracy, according to Proposition~\ref{prop:bound:accuracy}.\mer
\end{remark}

The proof of Theorem~\ref{thm:proposed:exact} requires the analysis of several properties, as performed in the next subsections.
The proof itself is given afterwards, in Section~\ref{sec:proofs:continuous}.

\subsection{Adaptation of the time-difference parameter}

To motivate the adaptation of $\hat T(t)$ in \eqref{eq:def:That:t}, the following lemma is obtained essentially as a corollary of Lemma~\ref{lem:diff:linear}.
\begin{lemma}
    \label{lem:diff:lineartimevarying}
    Let $N, L \in \RR_{\ge 0}$ and consider the differentiator $\Diff$ with output $y = \Diff u$ defined in \eqref{eq:proposed:diff}.
    Suppose that $f \in \FL$, $\eta \in \EN$, and $u = f + \eta$.
    Then,
    \begin{equation}
        \label{eq:accuracy:lineartimevarying}
        \abs{y(t) - \dot f(t)} \le \frac{2 N}{\hat T(t)} + \frac{L \hat T(t)}{2}
    \end{equation}
    holds whenever $t \ge \hat T(t) > 0$.
\end{lemma}
\begin{pf}
    Replace $T$ by $\hat T(t)$ in the proof of Lemma~\ref{lem:diff:linear}.
    \QED
\end{pf}

It is straightforward to verify that this lemma yields the desired optimal error bound \changed{$|y(t) - \dot f(t)| \le 2 \sqrt{2 N L}$}
if $\hat T(t) = 2\gamma(t) \sqrt{N/L}$   with $\gamma(t) \in [\sqrt{2}-1,1 +\sqrt{2}]$ for all $t$.
This fact motivates the structure of $\hat T(t)$ in \eqref{eq:def:That:t}.

Since $N$ is not available, an estimate $\hat N(t)$, given in \eqref{eq:def:Nhat:t}, is used instead of $N$ in the actual computation of $\hat T(t)$.
The properties of this noise amplitude estimate are analyzed next.
It will be shown that it is not necessarily equal to, but always bounded from above by $N$; this fact intuitively explains the restriction of $\gamma(t)$ to the interval $[1,\gammamax]$ with $\gammamax \le 1 + \sqrt{2}$ as opposed to also allowing values $\gamma(t) < 1$ as above.

\subsection{Properties of the noise estimate}

\changed{This section discusses upper and lower bounds on the noise estimate \eqref{eq:def:Nhat:t} and shows its relation to differentiability and growth bound of the measurements.}

\changed{\subsubsection{Upper and lower bounds for $\hat N$}}

To illuminate the noise amplitude estimation in \eqref{eq:def:Nhat:t}--\eqref{eq:def:aT:t} \changed{and to obtain an upper bound for $\hat N$}, the following lemma shows how $Q$ in \eqref{eq:def:aT:t} is related to the noise bound $N$.
\begin{lemma}
    \label{lem:aT}
    Let $L, N \in \RR_{\ge 0}$.
    For \changed{any} $t \in \RR_{> 0}$,  consider $Q(t,T,\sigma)$ as defined in \eqref{eq:def:aT:t} with $u \in \FL + \EN$.
    Then,
    \begin{equation}
        \abs{Q(t,T,\sigma)} \le 2N + \frac{L \sigma (T - \sigma)}{2}
    \end{equation}
    holds for all $T \in (0, t]$ and all $\sigma \in [0, T]$.
\end{lemma}
\begin{pf}
\changed{For arbitrary, fixed $t > 0$ and for any function $w \in \{ u, f, \eta \}$, define
\begin{equation}
    \label{eq:def:aT}
    a^w_T(\sigma) = w(t - \sigma) - w(t) + \frac{w(t) - w(t-T)}{T} \sigma.
\end{equation}
Then, $Q(t,T,\sigma) = a^u_T(\sigma) = a^\eta_T(\sigma) +a^f_T(\sigma)$.}
An upper bound for \changed{$a^\eta_T$} is given by
    \begin{align*}
        \abs{\changed{a^\eta_T}(\sigma)} &\le \abs{\eta(t- \sigma)} + \left(1 - \frac{\sigma}{T}\right) \abs{\eta(t)} + \frac{\sigma}{T} \abs{\eta(t-T)} \nonumber \\
        &\le N + \left(1 - \frac{\sigma}{T}\right) N + \frac{\sigma}{T} N = 2N.
    \end{align*}
    Moreover, $\changed{a^f_T}$ is continuously differentiable and satisfies
    \begin{equation*}
        \changed{a^f_T}(0) = \changed{a^f_T}(T) = 0, \qquad
        \changed{\ddot a^f_T}(\sigma) = \ddot f(t - \sigma),
    \end{equation*}
    and hence a.e. $\abs{\changed{\ddot a^f_T(\sigma)}}\le L$.
    Subject to these constraints, the extremal is given by
    \begin{equation*}
        \changed{\overline{a}^f_T}(\sigma) = \frac{L \sigma (T - \sigma)}{2}
    \end{equation*}
    i.e., \changed{$|a^f_T(\sigma)|\le \overline{a}^f_T(\sigma)$}.
The proof is concluded by noting that \changed{$|a^u_T(\sigma)|\le |a^\eta_T(\sigma)| + |a^f_T(\sigma)| \le 2N + \overline{a}^f_T(\sigma)$}.
    \QED
\end{pf}

Lemma~\ref{lem:aT} suggests that an estimate for $N$ may be obtained by subtracting the parabola arc $L \sigma (T -\sigma)/2$ from $|Q(t,T,\sigma)|$.
Indeed, the proposed estimate $\hat N(t)$ in \eqref{eq:def:Nhat:t} can be seen to be constructed by taking the supremum of this difference over all $0 < \sigma \le T \le \Tmax$.

Clearly $\hat N(t)$ is bounded by $N$ from above by construction; moreover, discontinuities in the noise impose a lower bound.
These properties are summarized in the following lemma.
\begin{prp}
\label{prop:Nhat:properties}
    Let $L, N \in \RR_{\ge 0}$, $\Tmax \in \RR_{> 0} \cup \{ \infty \}$, suppose $u = f + \eta$ with $f \in \FL$ and $\eta \in \EN$,  and consider a fixed $t \in \RR_{> 0}$.
    Define the (right-sided) discontinuity of $\eta$ at $t_0$ as
    \begin{equation}
        D(t_0) = \limsup_{\tau \to t_0^+} | \eta(\tau) - \eta(t_0)|.
    \end{equation}
    Then, $\hat N(t)$ as defined in \eqref{eq:def:Nhat:t} satisfies
    \begin{equation}
         \frac{D(t_0)}{2} \le \hat N(t) \le N
    \end{equation}
    for all $t_0 \in [0, t)$ with $t_0 \ge t-\Tmax$.
\end{prp}
\begin{pf}
The inequality $\hat N(t) \le N$ follows from the upper bound on $|Q(t,T,\sigma)|$ from Lemma~\ref{lem:aT}.
To show also $D(t_0) \le 2 \hat N(t)$,  fix $t_0 \in [0,t)$ with $t_0 \ge t-\Tmax$, and let $(\tau_k)$ be a sequence with $\tau_k \in [t_0,t)$, $\tau_k \to t_0$, and $|\eta(\tau_k) - \eta(t_0)| \to D(t_0)$.
\changed{Let $\sigma_k = t - \tau_k \ge 0$, $T = t - t_0 > 0$ and define $a^g_T$ as in \eqref{eq:def:aT} for $g \in \{u,f,\eta\}$, allowing to write $Q(t,T,\sigma) = a^\eta_T(\sigma) + a^f_T(\sigma)$.
    Then, $\sigma_k \to T \le \Tmax$ and due to the continuity of $f$ (and thus of $a^f_T$), one has}
    \begin{align*}
        \lim_{k \to \infty} \changed{a^f_{T}}(\sigma_k) = 0 = \lim_{k \to \infty} \frac{L \sigma_k (T - \sigma_k)}{2}.
    \end{align*}
    Hence, $2 \hat N(t)$ is bounded from below by
    \begin{align*}
        2 \hat N(t) &\ge \lim_{k \to \infty} \abs{\changed{a^\eta_{T}}(\sigma_k)} =  \lim_{k \to \infty} \abs{\eta(t - \sigma_k) - \eta(t-T)} \nonumber \\
        &= \lim_{k \to \infty} |\eta(\tau_k) - \eta(t_0)| = D(t_0),
    \end{align*}
    concluding the proof.
\QED
\end{pf}

Depending on whether $\hat N(t)$ at a given time instant $t > 0$ is zero or nonzero, two cases need to be distinguished.

\changed{\subsubsection{Differentiability and growth bound for $\hat N = 0$}}

If $\hat N(t)$ is zero, then the estimation (at time instant $t$) does not detect the presence of noise.
In this case, $Q$ in \eqref{eq:def:aT:t} satisfies
\begin{equation}
\label{eq:aTbound:noisefree}
    \abs{Q(t,T,\sigma)} \le \frac{L \sigma (T - \sigma)}{2}
\end{equation}
for $0 < \sigma \le T \le \Tmax$ by definition of $\hat N$ in \eqref{eq:def:Nhat:t}.
In the following, it will be shown that a (one-sided) derivative of the measurement $u$ exists in this case, and that $u$ satisfies a similar growth bound as $f$ in Lemma~\ref{lem:f:changerate}.

The following lemma shows that $\hat N(t) = 0$ implies left-sided differentiability of the noisy measurement $u$ at $t$.

\begin{lemma}
\label{lem:Nhat:zero:diff}
    Let $L \in \RR_{\ge 0}$, suppose that $u \in \mathcal{U}$, and consider a fixed $t \in \RR_{> 0}$.
    Let $\mu \in (0, t]$ and suppose $Q$ as defined in \eqref{eq:def:aT:t} satisfies \eqref{eq:aTbound:noisefree}
for all $T \in (0, \mu]$ and all $\sigma \in [0, T]$.
    Then, the limit
    \begin{equation}
    \label{eq:beta}
        \beta := \lim_{T \to 0^+} \frac{u(t) - u(t-T)}{T}
    \end{equation}
    exists.
\end{lemma}
\begin{pf}
    \changed{Consider the function $h : (0, \mu] \to \RR$ defined by
    \begin{equation}
    \label{eq:auxfunction2:h}
        h(\sigma) = \frac{u(t) - u(t-\sigma)}{\sigma}.
    \end{equation}
    From \eqref{eq:def:aT:t}, one obtains $Q(t,T,\sigma) = \sigma \bigl( h(\sigma) - h(T) \bigr)$.
    For given $\epsilon > 0$, consider arbitrary $\sigma, T$ satisfying the inequalities $\min(\epsilon,\mu) > T > \sigma > 0$.
    Then,
    \begin{equation}
        \abs{h(\sigma) - h(T)} = \frac{\abs{Q(t,T,\sigma)}}{\sigma} \le \frac{L (T-\sigma)}{2} < \frac{L \epsilon}{2}
    \end{equation}
    which implies existence of the limit $\beta = \lim_{\sigma  \to 0^+} h(\sigma)$.}
    \QED
\end{pf}

The next lemma shows that the measurements satisfy a similar growth bound as $f$ in Lemma~\ref{lem:f:changerate}.

\begin{lemma}
\label{lem:Nhat:zero:changerate}
Suppose that the conditions of Lemma~\ref{lem:Nhat:zero:diff} are fulfilled and let $\beta$ be defined as in \eqref{eq:beta} in that lemma.
     Then,
    \begin{equation}
\label{eq:f:parabola:beta}
    \abs{u(t - \sigma) + \beta \sigma - u(t)} \le \frac{L \sigma^2}{2}
    \end{equation}
    holds for all $\sigma \in [0, \mu]$.
\end{lemma}
\begin{pf}
    \changed{Consider the function $h : [0, \mu] \to \RR$ defined by \eqref{eq:auxfunction2:h} for $\sigma > 0$ and $h(0) = \beta$.
    According to Lemma~\ref{lem:Nhat:zero:diff}, this function is continuous at $\sigma = 0$ and by definition satisfies $u(t-\sigma) + \beta \sigma - u(t) = \sigma \bigl( h(0) - h(\sigma) \bigr)$.
    Using \eqref{eq:aTbound:noisefree}, one thus obtains
    \begin{align}
        \abs{h(0) - h(\sigma)} &= \abs{\sum_{i=0}^{\infty} h(\sigma/2^{i+1}) - h(\sigma/2^i)} \nonumber\\
        &\le \sum_{i=0}^{\infty} \frac{\abs{Q(t,\sigma/2^{i+1},\sigma/2^i)}}{\sigma/2^i} \nonumber \\
        &\le \sum_{i=0}^{\infty} \frac{L}{2} \bigl[ \sigma/2^{i} - \sigma/2^{i+1} \bigr] = \frac{L \sigma}{2}
    \end{align}
    for $\sigma \in [0,\mu]$, yielding the claimed inequality.}
    \QED
\end{pf}

\changed{\subsubsection{Growth bound for $\hat N > 0$}}

If $\hat N(t)$ is non-zero, then the measurements allow to distinguish noise, whose magnitude $N$ is at least $\hat N$.
In this case, it is neither possible nor necessary to compute an exact derivative of $u$ at time instant $t$.

The following lemma shows that a similar inequality as in Lemma~\ref{lem:Nhat:zero:changerate} may nonetheless be obtained with a suitable value of $\beta$.

\begin{lemma}
    \label{lem:auxlem}
    Let $L, \hat N \in \RR_{> 0}$, let $u \in \mathcal{U}$, and consider a fixed $t \in \RR_{> 0}$.
    Let $\hat T \ge 2 \sqrt{\hat N/L}$, define
\begin{equation}
        \beta := \frac{u(t) - u(t- \hat T)}{\hat T},\end{equation}
    and suppose that $Q$ as defined in \eqref{eq:def:aT:t} satisfies
    \begin{equation}
        \label{eq:auxlem:aT}
\abs{Q(t,\hat \sigma, \hat T)} \le 2 \hat N + \frac{L \hat T (\hat \sigma - \hat T)}{2}
    \end{equation}
for some $\hat \sigma \in [\hat T, t]$.
    Then,~\eqref{eq:f:parabola:beta} 
holds for $\sigma = \hat \sigma$.
\end{lemma}
\begin{remark}
    Note that for $\hat N = \hat N(t)$ as defined in \eqref{eq:def:Nhat:t}, condition \eqref{eq:auxlem:aT} of this lemma is fulfilled for every $\hat \sigma \in [\hat T, t]$ satisfying $\hat \sigma \le \Tmax$.\mer
\end{remark}
\begin{pf}
     \changed{Consider the function $h : (0, t] \to \RR$ defined by \eqref{eq:auxfunction2:h}.
     Then, $\beta = h(\hat T)$ and
     \begin{align}
         \frac{\abs{u(t - \hat\sigma) + \beta \hat\sigma - u(t)}}{\hat \sigma} &= \abs{h(\hat T) - h(\hat\sigma)} \le \frac{Q(t,\hat\sigma,\hat T)}{\hat T} \nonumber\\
         &\le \frac{2\hat N}{\hat T} + \frac{L (\hat \sigma - \hat T)}{2} \le \frac{L \hat \sigma}{2}
     \end{align}
     since $\hat T^2 \ge 4 \hat N/L$ implies $2\hat N/\hat T \le L \hat T/2$.}
     \QED
\end{pf}

\subsection{Worst-case error upper bound}
\label{sec:ored:accuracy}

To prove the optimal accuracy, the following lemma shows how a bound for the differentiation error may be obtained from the growth bounds proven in \changed{Lemmata}~\ref{lem:Nhat:zero:changerate} and~\ref{lem:auxlem}.
\begin{lemma}
\label{lem:acc:optimal}
    Let $L \in \RR_{> 0}$, $N \in \RR_{> 0}$ and $\beta \in \RR$.
    Define $\ell := \sqrt{2 N/L}$, let $t \ge \ell$ and suppose that $u = f + \eta$ with $f \in \FL$ and $\eta \in \EN$ satisfies
    \begin{equation}
        \label{eq:acc:bound:prev}
        |u(t - \sigma) - u(t) + \beta \sigma| \le \frac{L \sigma^2}{2}
    \end{equation}
    for $\sigma = \ell + x$ with $x \in [0,\Delta]$ for some $\Delta \in [0,\ell]$.
    Then, $|\beta - \dot f(t)| \le 2 \sqrt{2 N L} + L \dfrac{\Delta}{2}$.
\end{lemma}
\begin{remark}
The degree of freedom $\Delta$ introduced in this lemma is zero in the continuous-time case, which yields the desired optimal bound (cf. Proposition~\ref{prop:bound:exact}).
Later, $\Delta$ will be used to derive an error bound also for a discrete-time implementation of the differentiator.\mer
\end{remark}
\begin{pf}
According to Lemma~\ref{lem:f:changerate}, $f$ satisfies the inequality~\eqref{eq:f:der:growth}
for all $\sigma \in [0, t]$.
From \eqref{eq:acc:bound:prev}, one moreover has
$
    |f(t - \sigma) - f(t) + \beta \sigma| \le L \sigma^2/2 + 2 N
$
for $\sigma = \ell + x$ due to the fact that $f = u - \eta$ and $|\eta(t)|\le N$.
By combining the inequalities and dividing by $\sigma$, one obtains
\begin{equation}
    \label{eq:acc-bnd:sigma}
    \abs{\beta - \dot f(t)} \le L \sigma + \frac{2 N}{\sigma} 
\end{equation}
for $\sigma = \ell+x$. Next, the right-hand side of \eqref{eq:acc-bnd:sigma} is shown to be less than or equal to $2\sqrt{2NL} + L\Delta/2$. Since $x\in [0,\Delta]$ and $\Delta \in [0,\ell]$, it follows that
\begin{align*}
    0 &\ge (x+\Delta/2)(x-\Delta) = x^2 - \frac{\Delta}{2} x - \frac{\Delta^2}{2} \\
    &\ge x^2 - \frac{\Delta}{2} x - \frac{\Delta}{2}\ell = \sigma^2 - 2 \ell \sigma + \ell^2 - \frac{\Delta}{2} \sigma.
\end{align*}
Adding $2 \ell \sigma + \Delta \sigma/2$ and multiplying by $L/\sigma$
then yields
\begin{align*}
    2L \ell + L \frac{\Delta}{2} \ge L \sigma + L \frac{\ell^2}{\sigma}
\end{align*}
which gives
$2 \sqrt{2NL} + L \Delta/2 \ge L\sigma + 2N / \sigma$
by definition of $\ell$.
Combining this with \eqref{eq:acc-bnd:sigma} establishes the result.
\QED
\end{pf}

Using \changed{Lemmata}~\ref{lem:diff:lineartimevarying} and~\ref{lem:acc:optimal}, an upper bound for the worst-case differentiation error is now proven.
\begin{prp}
\label{thm:proposed:error}
Let $L, \epsilon > 0$ and consider the differentiator $\Diff$ defined by \eqref{eq:proposed:diff} with parameters $\gammamax \ge 1$ and $\Tmax \in \RR_{> 0} \cup \{ \infty \}$.
Let $\Nmax = L \Tmax^2/2$ and define the function $\Tc : \RR_{\ge 0} \times [0,\Nmax) \to \RR_{> 0}$ as
\begin{equation}
    \Tc(R, N) = \max \left\{\sqrt{2 N/L}, \epsilon\right\}.
\end{equation}
Then, $M_N^{L,R}[\Tc(R,N)] \le \max\{2 \sqrt{2}, \gammamax + \gammamax^{-1}\} \sqrt{ N L}$ holds for all $N \in (0, \Nmax)$ and all $R \ge 0$.
\end{prp}
\begin{pf}
    Let $N \in (0,\Nmax)$, $R \in \RR_{\ge 0}$ and consider a fixed positive $t \ge \Tc(R,N) \ge \sqrt{2N/L}$.
    Relation \eqref{eq:def:That:t} along with $\hat N(t) \le N$ and $\gamma(t) \le \gammamax$ imply $\hat T(t) \le 2 \gammamax \sqrt{N/L}$.
    Distinguish hence the cases $\hat T(t) \in [\sqrt{2 N/L}, 2 \gammamax \sqrt{N/L}]$ and $\hat T(t) \in [0, \sqrt{2 N/L})$.

    In the first case, according to Lemma~\ref{lem:diff:lineartimevarying},
    \begin{align*}
        \abs{y(t) - \dot f(t)} &\le \frac{2N}{\hat T(t)} + \frac{L \hat T(t)}{2} \nonumber\\
        &\le \max\Bigl\{\frac{2}{\sqrt{2}}+\frac{\sqrt{2}}{2}, \frac{1}{\gammamax} + \gammamax\Bigr\} \sqrt{N L} \end{align*}
    then holds, wherein $\sqrt{2} + \sqrt{2}^{-1} < 2 \sqrt{2}$.

    In the second case, $\hat T(t) \ge 2 \sqrt{\hat N(t)/L}$ follows from \eqref{eq:def:That:t}, because $\hat T(t) < \sqrt{2 N/L} \le t$ and hence both $\hat T(t) = t$ and $\hat T(t) = \Tmax > \sqrt{2 N/L}$ are impossible, and $\gamma(t) \ge 1$.
    Now distinguish the cases $\hat T(t) > 0$ and $\hat T(t) = 0$.
    For $\hat T(t) > 0$, use Lemma~\ref{lem:auxlem} to find that \eqref{eq:acc:bound:prev} holds for $\beta = y(t)$ and $\sigma = \sqrt{2 N/L}$, because $\Tmax \ge \sigma \ge  \hat T(t)$.
    For $\hat T(t) = 0$, conclude that also $\hat N(t) = 0$ and deduce from Lemma~\ref{lem:Nhat:zero:changerate} with $\mu = \min\{t,\Tmax\}$ that \eqref{eq:acc:bound:prev} holds for $\beta = y(t)$ and all $\sigma \in [0,\sqrt{2N/L}]$.
    The bound $|y(t) - \dot f(t)| \le 2 \sqrt{2 N L}$ then follows from Lemma~\ref{lem:acc:optimal} setting $\Delta = 0$.
    \QED
\end{pf}

\subsection{Proof of the Main Theorem~\ref{thm:proposed:exact}}
\label{sec:proofs:continuous}

\textbf{Proof of Theorem~\ref{thm:proposed:exact}\ref{item:welldef})} It is clear that the differentiator is well-defined if $\hat T(t) > 0$. If $\hat T(t) = 0$, then also $\hat N(t) = 0$ and hence $Q$ as defined in \eqref{eq:def:aT:t} satisfies \eqref{eq:aTbound:noisefree}.
Existence of the limit in~\eqref{eq:proposed:diff:y} for all $u \in \mathcal{U}$ then follows from Lemma~\ref{lem:Nhat:zero:diff}.

\textbf{Proof of Theorem~\ref{thm:proposed:exact}\ref{item:exact})} If $N = 0$, then $\hat T(t) = 0$ due to Proposition~\ref{prop:Nhat:properties}.
    Thus, \eqref{eq:proposed:diff:y} with $u = f$ implies $y(t) = \dot f(t)$, i.e., $M_0^{L,R}(t) = 0$, for all $t > 0$, proving exactness from the beginning over $\FL$.
    For robustness, fix $t > 0$ and set $\epsilon = t$ in Proposition~\ref{thm:proposed:error}.
    For all $N \in (0, L t^2/8]$ with $N \le L \Tmax^2/2$, then  $\Tc(R,N) = t$ in that proposition, and hence \begin{align*}
        \sup_{f \in \FL^R} Q_N^f(t) &\le M_N^{L,R}(t) + M_0^{L,R}(t) = M_N^{L,R}(t) \nonumber\\
        &\le \max\{2 \sqrt{2}, \gammamax+\gammamax^{-1}\} \sqrt{N L},
    \end{align*}
    which implies $Q^{L,R}(t) = \lim_{N \to 0}\sup_{f \in \FL^R} Q_N^f(t) = 0$, proving robustness almost from the beginning over $\FL$.

\textbf{Proof of Theorem~\ref{thm:proposed:exact}\ref{item:errorbound})}
Note that $\gammamax + \gammamax^{-1} \le 2 \sqrt{2}$ for $\gammamax \in [1,1+\sqrt{2}]$.
    Then, for every $t > \sqrt{2 N /L}$ find that $M_N^{L,R}(t) \le 2\sqrt{2 N L}$ for all $N \in (0, L \Tmax^2/4)$ due to Proposition~\ref{thm:proposed:error} with $\epsilon = t$ as well as for $N = 0$ due to Theorem~\ref{thm:proposed:exact}\ref{item:exact}).
    Consequently, $\CLa \le 2 \sqrt{2}$.
    Equality follows from the fact that $\CLa \ge 2 \sqrt{2}$ for all causal differentiators according to Proposition~\ref{prop:bound:accuracy}.

\textbf{Proof of Theorem~\ref{thm:proposed:exact}\ref{item:global})}
Theorem~\ref{thm:proposed:exact}\ref{item:errorbound}) with $\Tmax = \infty$ yields $M_N^{L,R}(t) \le 2\sqrt{2 N L}$ for all $N \in [0,\infty)$; hence, $\CLg \le 2 \sqrt{2}$, with equality again due to Proposition~\ref{prop:bound:accuracy}.
\QED

\section{Sample-Based Differentiation}
\label{sec:sampled-performance}

When only sampled information is available, the worst-case error between any estimate and the true derivative can never be better than if measurements are available continuously over time. Therefore, some degree of error additional to that obtained in the continuous-time case is to be expected. 
Taking this fact into account, it will be shown in this section that sampled versions of the proposed differentiators achieve best possible accuracy. Differentiators operating on signal samples will henceforth be called \emph{sample-based differentiators}.

In the following, suppose that only sampled measurements of $u = f + \eta$ are available, at times $t_k=k\Delta$, where $\Delta > 0$ is the sampling period and $k\in \NN_0$.
Such a differentiator, which takes the samples $u(t_j)$ from $j=0$ to $j=k$ as input to produce the estimate $y_k$ of the derivative $\dot{f}(t_k)$ is hereafter denoted by $\Diff_\Delta$, so that\footnote{Formally, a sample-based differentiator, $\Diff_{\Delta}$ with sampling time $\Delta > 0$ is an operator $\Diff_{\Delta} : \mathcal U \to (\Delta\cdot\NN_0 \to \RR)$, with the additional property that $\Diff_{\Delta}(u_1+u_2)$ = $\Diff_{\Delta}(u_1)$ whenever $u_2(k\Delta) = 0$ for all $k \in \NN_0$.} $[\Diff_\Delta u](t_k) = y_k$.

\subsection{Worst-case error}

The worst-case error in Definition~\ref{def:acc:abs} can be straightforwardly adapted to the sampled case, as follows.
\begin{defin}
    \label{def:acc:abs:sampled}
    Let $L, N \in \RR_{\ge 0}$ and $\Delta > 0$. A sample-based differentiator $\Diff_\Delta$ is said to have worst-case error $M_N^{L,R}(t)$ from time $t=k\Delta$, $k\in\NN_0$, over the signal class $\FL^R$ with noise bound $N$ if
            \begin{equation*}
                M_N^{L,R}(k\Delta) = \sup_{\substack{u=f+\eta \\ \eta\in\EN\\ f\in\FL^R}} \sup_{\substack{\ell\in\NN_0\\ \ell\ge k}} \big|\dot f(\ell\Delta) - [\Diff_\Delta u](\ell\Delta)\big|.
            \end{equation*}
\end{defin}
It is worth noting that for a sample-based differentiator $\Diff_\Delta$ the worst-case error $M_N^{L,R}$ is also only defined at integer multiples of the sampling time $\Delta$.
As in the continuous-time case, $M_N^{L,R}(k\Delta)$ is non-increasing with respect to $k \in \NN_0$ and non-decreasing with respect to $N, L, R \in \RR_{\ge 0}$.

\subsection{Performance limits and quasi-exactness}

\begin{figure}
    \centering
    \includegraphics{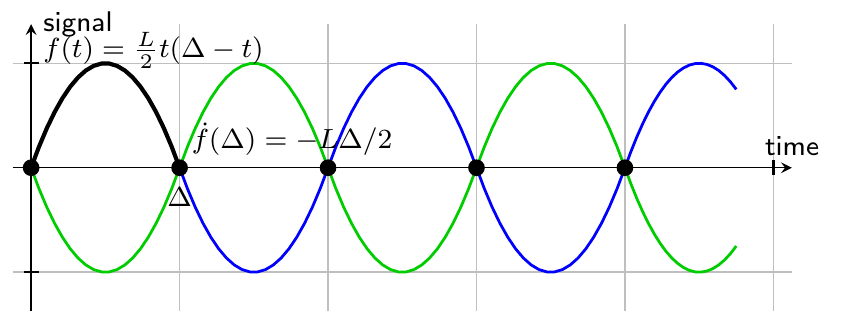}
    \caption{Black: parabola arc satisfying $\ddot f(t)\equiv -L$, $f(0)=0$ and $f(\Delta)=0$. Blue and green: signals constructed by piecing together shifted (and sign-changed) copies of the black parabola arc. Black circles: measurements (all zero).}
    \label{fig:DT_accuracy_no_noise}
\end{figure}

A lower bound on the additional error introduced by sampling is given in the following result.
\begin{prp}
    \label{prop:worst-case-ge-noise-free}
    Let $L, N, R \in \RR_{\ge 0}$, $\Delta > 0$. Then, the worst-case error of any sample-based differentiator $\Diff_\Delta$ satisfies
    \begin{equation}
    \label{eq:sampled-worst-case-lower-bound}
         M_N^{L,R}(k\Delta) \ge \frac{L \Delta}{2},\quad\forall k\in\NN.
    \end{equation}
\end{prp}
\begin{remark}
An illustration of the proof is shown in Fig.~\ref{fig:DT_accuracy_no_noise}.
Essentially, piecing together parabola arcs with second derivative alternately equal to $\pm L$ on time intervals of length $\Delta$ yields zero measurements, making two functions with maximum derivative $\pm L \Delta/2$ indistinguishable to the differentiator from the sampled measurements alone.\mer
\end{remark}
\begin{pf}
    The fact that $M_N^{L,R}(k\Delta) \ge M_0^{L,0}(k\Delta)$ follows directly from the definition, hence it suffices to show $M_0^{L,0}(k\Delta) \ge \frac{L \Delta}{2}$.
    Assume to the contrary that there exist $k_0 \in \NN$ and $\epsilon > 0$ such that $M_0^{L,0}(k\Delta) < (1-\epsilon) \frac{L \Delta}{2}$ for $k = k_0$.
    By Definition~\ref{def:acc:abs:sampled}, this is then true also for all integers $k \ge k_0$.

    Consider the function $g_{a,b,c}: [0,\Delta] \to \RR$ with parameters $a, b\in [0,1]$ and $c \in [0, \frac{1}{4}]$, defined as
    \begin{equation*}
        g_{a,b,c}(t) = \begin{cases}
            a \frac{L \Delta}{2} t + \frac{L t^2}{2} & t \in [0, c \Delta) \\
            b \frac{L \Delta^2}{8} - \frac{L}{2} (t - \frac{\Delta}{2})^2 & t \in [c \Delta, \frac{\Delta}{2}) \\
            b \frac{L}{2} t (\Delta - t) & t \in [\frac{\Delta}{2}, \Delta].
        \end{cases}
    \end{equation*}
    For arbitrary $a \in [0, 1]$, $b = 1 - \frac{1}{2} (1 - a)^2$, and $c = \frac{1}{4} (1 - a)$,
    one verifies that $|\ddot g_{a,b,c}(t)| \le L$ almost everywhere on $[0, \Delta]$, that $g_{a,b,c}(0) = g_{a,b,c}(\Delta) = 0$, that $\dot g_{a,b,c}(0) = a \frac{L \Delta}{2}$, and that $\dot g_{a,b,c}(\Delta) = - b \frac{L \Delta}{2}$.
    In particular, $g_{1,1,0}(t) = \frac{L}{2} t (\Delta-t)$ is the black parabola arc depicted in Fig.~\ref{fig:DT_accuracy_no_noise}.

    Recursively define the (strictly increasing) sequence $(a_j)$ via $a_{j+1} = 1 - \frac{1}{2} (1 - a_j)^2$ with $a_0 = 0$, and set $c_{j} = (1-a_j)/4$, $b_j = a_{j+1}$ for all $j \in \NN_0$.
    Using these sequences, define the function $f : \RR_{\ge 0} \to \RR$ piece-wise as
    \begin{equation*}
        f(t) = (-1)^j g_{a_j,b_j,c_j}(t-j\Delta) \qquad \text{for $t \in [j \Delta, (j+1) \Delta)$.}
    \end{equation*}
    From the properties of $g_{a,b,c}$ above, it follows that $f \in \FL^0$ with $f(k\Delta) = 0$ for all $k \in \NN_0$.
    Applying the inputs $u = f$ and $u = -f$ to the differentiator then produces identical (zero) measurements at the sampling time instants.
    Hence, $M_0^{L,0}(k \Delta) \ge |\dot f(k\Delta)| = a_k \frac{L \Delta}{2}$.
    Since $\lim_{j\to \infty} a_j = 1$, it is possible to select $k \ge k_0$ such that $a_k \ge 1-\epsilon$, yielding a contradiction to $M_0^{L,0}(k\Delta) < (1-\epsilon) \frac{L \Delta}{2}$.
    \QED
\end{pf}

As the previous result shows, it is clearly impossible to obtain an exact differentiator based on sampled measurements.
Arguably, the closest property to exactness in the sampled case is to achieve equality in \eqref{eq:sampled-worst-case-lower-bound}.
This property is called quasi-exactness in the following.
Its formal definition is similar to that of exactness in Definition~\ref{def:exact}.

\begin{defin}
\label{def:quasi-exact}
    A sample-based differentiator $\Diff_{\Delta}$ is said, over the signal class $\FL$, to be
    \begin{itemize}
        \item
            quasi-exact in finite time if for every $R$ there exists $k_R \in \NN$ such that $M_0^{L,R}(k_R\Delta) = \frac{L \Delta}{2}$\changed{;}
        \item
            quasi-exact in fixed time if there exists $k \in \NN$ such that $M_0^{L,R}(k\Delta) = \frac{L \Delta}{2}$ for all $R \ge 0$\changed{;}
        \item
            quasi-exact from the beginning if $M_0^{L,R}(\Delta) = \frac{L \Delta}{2}$ for all $R \ge 0$\changed{;}
        \item
            not quasi-exact if it is not quasi-exact in finite time.
    \end{itemize}
\end{defin}
A quasi-exact sample-based differentiator yields the best possible estimate for the derivative in the noise-free case based on the available samples.
It is worth noting, however, that the discretization of an exact continuous-time differentiator does not necessarily yield a quasi-exact sample-based differentiator.

The simplest quasi-exact differentiator is the first-order difference $y_k = [u(t_k) - u(t_k-\Delta)]/\Delta$, as can be seen from Lemma~\ref{lem:diff:linear} setting $N = 0$.
In the presence of noise, however, the worst-case error of this simple differentiator may become prohibitively large.
It is hence desirable to achieve a worst-case error that is close to the lower bound of all causal quasi-exact differentiators, which is stated in the following proposition that is analogous to Proposition~\ref{prop:bound:exact}.

\begin{prp}
\label{prop:sampled-lower-bound}
Let $\Delta > 0$, $L \ge 0$ and consider a causal sample-based differentiator $\Diff_\Delta$.
Suppose that $\Diff_{\Delta}$ is quasi-exact in finite-time over $\FL$.
Then,
\begin{equation}
\label{eq:sampled-quasi-exact-lower-bound}
    M_N^{L,R}(r\Delta) \ge 2 \sqrt{2 N L} - \frac{L \Delta}{2}
\end{equation}
holds for all $R, N \ge 0$ and $r\in\NN$. \end{prp}
\begin{remark}
Sample-based differentiators are trivially limited also by the bound $M_N^{L,R}(k\Delta) \ge 2 \sqrt{N L}$ from Proposition~\ref{prop:bound:causal}, which applies to every causal differentiator.
Hence, the bound \eqref{eq:sampled-quasi-exact-lower-bound} is nontrivial only if $\Delta \le 4 (\sqrt{2} - 1) \sqrt{N/L}$.\mer
\end{remark}
\begin{pf}
The statement is trivial for $L = 0$.
For $L > 0$, let $k \in \NN$ be such that $M_0^{L,0}(k\Delta) = L \Delta/2$ according to Definition~\ref{def:quasi-exact}.
Furthermore, let $N\ge 0$, $r\in\NN$, and define $\kappa = \sqrt{N/L}$, select $\ell\in\NN$ such that $\ell\ge \max\{r,k\}$ and $\ell\Delta \ge (2+\sqrt{2})\kappa$. Define also
$T=\ell\Delta$ and $\tau = \ell\Delta - (2+\sqrt{2})\kappa$.
For these values of $\kappa, \tau, T$, consider the functions $g_1$ and $g_2$ in \eqref{eq:g1-exact}--\eqref{eq:g2-exact}.
Then, $g_1 \in \FL^0$ and $g_2 \in \EN$.
    Choosing $f = -g_1$ and $\eta = g_2$ yields differentiator input $u(t) = g_1(t)$ for all $t \in [0, T]$.
    Since $g_1 \in \FL^0$, then
    \begin{align}
    \frac{L \Delta}{2} &\ge M_0^{L,0}(k\Delta) \ge M_0^{L,0}(T) \ge 
        |[\Diff_{\Delta} u](T) - \dot g_1(T)| \nonumber \\
        &\ge 2 |\dot g_1(T)| - |[\Diff_{\Delta} u](T) + \dot g_1(T)|.
    \end{align}
    Since $\dot f(T) = -\dot g_1(T) = - \sqrt{2 N L}$, and $M_N^{L,R}(j\Delta)$ is non-decreasing with respect to $R$ and non-increasing with respect to $j$, this yields
    \begin{align*}
        M_N^{L,R}(r\Delta) &\ge M_N^{L,R}(T) \ge |[\Diff_{\Delta} u](T) - \dot f(T)| \nonumber \\
        &= |[\Diff_{\Delta} u](T) + \dot g_1(T)| \ge 2 \sqrt{2 N L} - \frac{L \Delta}{2},
    \end{align*}
    as claimed.
    \QED
\end{pf}

\section{Sample-Based Optimal Robust Exact Differentiation}
\label{sec:sampled-adaptive-lin-diff}

In this section, a sampled version of the proposed optimal robust exact differentiator is shown.

\subsection{Proposed sample-based differentiator}
\label{sec:sample-based-diff}

The sampled version of \eqref{eq:proposed:diff} is given by the noise amplitude estimation over a time window of length $\Tmax = \kmax \Delta$ with parameter\footnote{Like in the continuous-time case, all formal results also hold with infinite window length, i.e., with $\kmax = \infty$.} $\kmax \in \NN \setminus \{ 1 \}$ according to \changed{$\hat N_0 = \hat N_1 = 0$ and}
\begin{subequations}
\label{eq:diff:sampled}
\begin{align}
\label{eq:diff:sampled:Nhat}
\hat N_k &= \frac{1}{2} \max_{\substack{\ell \in \{ 2, \ldots, \kmax \} \\ \ell \le k \\ j \in \{1, \ldots, \ell\}}} \left(\abs{Q(t_k, \ell \Delta, j \Delta)} - \frac{L \Delta^2 j (\ell -j)}{2}\right), \\
\intertext{\changed{for $k \ge 2$,} the selection of $\gamma_k$ according to\footnotemark}
\noalign{\footnotetext{\changed{In practice, the smallest element of $\mathcal{G}_k \cap [1,\gammamax]$ may be chosen.}}}
\gamma_k &\in \mathcal{G}_k \cap [1, \gammamax],\quad\text{with }\gammamax \ge 2 \text{ and} \\
\mathcal{G}_k &= \begin{cases}
        \{ 1 \} & \text{if }2 \sqrt{ \frac{\hat N_k}{L}} \le \Delta, \\
        \left\{ \frac{j \Delta}{2 \sqrt{\hat N_k /L}} : j \in \NN \right\} & \text{otherwise,}
        \end{cases} \\
\intertext{and the computation of the differentiator output via}
\label{eq:hatTk:sampled}
\hat T_k &= \min\left\{ t_k, \kmax \Delta, \max\left\{\Delta, 2 \gamma_k \sqrt{\frac{\hat N_k}{L}}\right\}\right\} \\
y_k &= \frac{u(t_k) - u(t_k - \hat T_k)}{\max\{\hat T_k,\Delta\}}
\end{align}
\end{subequations}
with $Q$ in~\eqref{eq:def:aT:t}. 
As will be shown, the restriction $\gammamax \ge 2$ ensures that the set $\mathcal G_k \cap [1,\gammamax]$ is always non-empty and that, for $k \ge 1$, $\hat T_k / \Delta  \in \NN$ and $\hat T_k = \Delta$ if and only if $\mathcal{G}_k = \{ 1 \}$.
From Lemma~\ref{lem:aT}, it follows that $\hat N_k \le N$ in analogy to Proposition~\ref{prop:Nhat:properties}.

It will be shown that \changed{the} sample-based implementation \changed{\eqref{eq:diff:sampled}} of the optimal robust exact differentiator in \eqref{eq:proposed:diff} is quasi-exact, and that, with appropriately chosen $\gammamax$, its worst-case differentiation error $M_{N}^{L,R}(k\Delta)$ is always contained in a band $2\sqrt{2 N L} \pm L \Delta/2$ around its optimal continuous-time value.
Considering Propositions~\ref{prop:worst-case-ge-noise-free} and~\ref{prop:sampled-lower-bound}, one can see that this is the tightest band of this form one can hope to obtain for a quasi-exact sample-based differentiator.
The following main theorem for the sampled case formally states these results.
The proof is given in Section~\ref{sec:proofs:sampled}.
\begin{theorem}
    \label{thm:proposed:quasi-exact}
    \label{thm:sampled:errorband}
    Let $L, \Delta \in \RR_{> 0}$,  $N \in \RR_{\ge 0}$, and consider the sample-based differentiator $\Diff_\Delta$ as given in \eqref{eq:diff:sampled} with parameters $\gammamax \ge 2$ and $\kmax \in \NN \setminus \{ 1 \}$. Let $\Nmax = L \Delta^2 (\kmax-1)^2/2$.
    Then, \changed{the following statements are true:}
    \begin{enumerate}[a)]
        \item $\Diff_{\Delta}$ is a well-defined sample-based differentiator; specifically, $y_k = 0$ for $k = 0$, and, for all $k \in \NN$, the set $\mathcal{G}_k \cap [1, \gammamax]$ is non-empty and $\hat T_k/\Delta \in \NN$.\label{item:non-empty}
        \item $\Diff_{\Delta}$ is quasi-exact from the beginning over $\FL$.\label{item:quasi-exact}
        \item If $\gammamax \in [2, 1+\sqrt{2}]$, then     
        \begin{equation*}
            2 \sqrt{2 N L} - \frac{L \Delta}{2} \le M_N^{L,R}(k\Delta) \le  2 \sqrt{2 N L} + \frac{L \Delta}{2}
        \end{equation*}
    for all $N \le \Nmax$ and for all $k \in \NN$ with $k\Delta \ge 2\sqrt{N/L}$.
    In particular, $M_N^{L,R}(k\Delta) = 2 \sqrt{2 N L} + \frac{L \Delta}{2}$ for $N = 0$ and all $k \in \NN$.
\label{item:errorband}
    \item If $\gammamax \in [2,3/\sqrt{2})$, then $M_N^{L,R}(k\Delta) = 2 \sqrt{2 N L} - \frac{L \Delta}{2}$ for $N = L\Delta^2/2$ and all $k \in \NN$ with $k \ge 2$.\label{item:sampled:lower} \met
    \end{enumerate}
\end{theorem}

\changed{\begin{remark}[Tuning]
Analogous to Remark~\ref{rem:tuning}, the proposed sample-based differentiator may be tuned using a (crude) noise amplitude upper bound $\Nmax$ by selecting the smallest integer $\kmax$ satisfying $\kmax\Delta > \sqrt{2 \Nmax /L} +\Delta$, setting $\gammamax = 2$, and choosing the smallest possible value for $\gamma_k$ in every sampling step.\mer
\end{remark}}
\begin{remark}
The theorem reveals that, \changed{for fixed $L$}, the upper differentiation error bound has the same asymptotic behavior \changed{of the order $\max\{ \sqrt{N}, \Delta \}$} with respect to sampling period $\Delta$ and noise amplitude $N$ as existing robust exact differentiators with sampled measurements, cf. \cite[Theorems 1 and 3]{levant2014proper}.\mer
\end{remark}

\newcommand{\Kc}{\mathcal{K}}
\subsection{Worst-case error upper bound}
The following result gives an upper bound on the worst-case error similar to the one in Proposition~\ref{thm:proposed:error}.
\begin{prp}
    \label{thm:error:sampled:tight}
    Let $L, \Delta \in \RR_{> 0}$, $u = f + \eta$ with $f \in \FL$, $\eta \in \EN$, and consider the sample-based differentiator $\Diff_\Delta$ as given in \eqref{eq:diff:sampled} with parameters $\gammamax \ge 2$ and $\kmax \in \NN \setminus \{ 1 \}$.
    Define $\Nmax = L \Delta^2 (\kmax-1)^2/2$.
    Then, the worst-case error bound $M_N^{L,R}(k \Delta) \le \max\{2\sqrt{2}, \gammamax + \gammamax^{-1}\} \sqrt{ N L} + L \frac{\Delta}{2}$ holds for all $N \in [0, \Nmax]$ and all $k \in \NN$ with $k \Delta \ge \sqrt{2 N/L}$.
\end{prp}
\begin{pf}
Let $t_k = k \Delta$, $N \in [0,\Nmax]$ and distinguish cases $N \le L \hat T_k^2/2$ and $N > L \hat T_k^2/2$.
    In the first case, $\hat T_k \ge \sqrt{2 N/L}$ holds.
    With the inequality $\hat N_k \le N$, which follows from the definition of $\hat N_k$ and Lemma~\ref{lem:aT}, moreover either $\hat T_k \le 2 \gamma_k \sqrt{N/L}$ or $\hat T_k = \Delta$ holds.
    According to Lemma~\ref{lem:diff:lineartimevarying},
    \begin{align}
        \abs{y_k - \dot f(t_k)} &\le \frac{2N}{\hat T_k} + \frac{L \hat T_k}{2} \nonumber\\
        \intertext{and if $\hat T_k \le 2 \gamma_k \sqrt{N/L}$, then}
        \frac{2N}{\hat T_k} + \frac{L \hat T_k}{2} &\le \max\Bigl\{\frac{2}{\sqrt{2}}+\frac{\sqrt{2}}{2}, \frac{1}{\gamma_k} + \gamma_k\Bigr\} \sqrt{N L} \nonumber\\
&\le \max\{ 2\sqrt{2}, \gammamax + \gammamax^{-1}\} \sqrt{N L}.
    \end{align}
    holds. If $\hat T_k = \Delta$, then
    \begin{align}
        \label{eq:inter-bound}
        \frac{2N}{\hat T_k} + \frac{L \hat T_k}{2} &\le \frac{2N}{\sqrt{2N/L}} + L\frac{\Delta}{2} = \sqrt{2NL} + L\frac{\Delta}{2}.
    \end{align}
    In the second case, we have $\hat T_k < \sqrt{2N/L} \le t_k$.
    Then, $\hat T_k \ge 2\sqrt{\hat N_k / L}$ follows from \eqref{eq:hatTk:sampled}, because $\hat T_k = t_k$ and $\hat T_k = \kmax \Delta$ are impossible due to $(\kmax - 1) \Delta \ge \sqrt{2N/L}$, because $\hat T_k = \Delta$ implies $\mathcal{G}_k = \{ 1 \}$ and $\Delta \ge 2 \sqrt{\hat N_k/L}$, and because $\gamma_k \ge 1$.
    Define $\ell = \sqrt{2N/L}$ and
    let $x\in [0,\Delta)$ be such that $\hat\sigma := \ell + x$ satisfies $\hat\sigma/\Delta \in \NN$.
    Then, moreover $\hat \sigma \le \sqrt{2 N/L} +\Delta \le \kmax \Delta$, and by definition of $\hat N_k$ in \eqref{eq:diff:sampled:Nhat},
    \begin{align*}
        \abs{Q(t_k,\hat \sigma,\hat T_k)}  \le 2\hat N_k + \frac{L\hat T_k(\hat\sigma - \hat T_k)}{2}
    \end{align*}
    holds because $\hat T_k \le \hat\sigma \le t_k$ and $\hat\sigma/\Delta \in \NN \cap [1,\kmax]$.
    Using Lemma~\ref{lem:auxlem}, then \eqref{eq:f:parabola:beta} holds with $t=t_k$ and $\sigma = \hat\sigma$. The result then follows from Lemma~\ref{lem:acc:optimal}.\QED
\end{pf}

\subsection{Worst-case error lower bound}

Proposition~\ref{thm:error:sampled:tight} shows that for $N = 0$, the worst-case error bound is equal to $L \Delta /2$, corresponding to the noise-free case. This noise-free bound is tight and cannot be improved, as is shown in Proposition~\ref{prop:worst-case-ge-noise-free}. 

For all $N \ge 0$, a lower bound on the worst-case error is given by Proposition~\ref{prop:sampled-lower-bound}.
The following auxiliary lemma will be used to show that the proposed sample-based differentiator can actually attain this lower bound in some specific cases.
\begin{lemma}
    \label{lem:better-than-conttime}
    Let $L \in \RR_{\ge 0}$, $\Delta > 0$ and $N=L\Delta^2/2$. Then, the sample-based differentiator $\Diff_\Delta$ given in \eqref{eq:diff:sampled} with parameters $\gammamax \ge 2$ and $\kmax \in \NN \setminus \{1\}$ achieves
    \[  M_N^{L,R}(t_k) \le 2 \sqrt{2 N L} - \frac{L \Delta}{2} \]    
    whenever $t_k \ge \hat T_k$ and $\hat T_k \in \{\Delta,2\Delta\}$.
\end{lemma}
\begin{pf}
According to Lemma~\ref{lem:diff:lineartimevarying} and provided $t_k \ge \hat T_k$, it follows that
\begin{align*}
    \abs{[\Diff_\Delta u](t_k) - \dot{f}(t_k)} &\le \frac{2 N}{\hat T_k} + \frac{L \hat T_k}{2} = \frac{3}{2} L \Delta \nonumber \\
    &= 2 \sqrt{2 N L} - \frac{L \Delta}{2}
\end{align*}
for the considered value $N = L \Delta^2/2$.
\QED
\end{pf}

\subsection{Proof of the Main Theorem~\ref{thm:proposed:quasi-exact}}
\label{sec:proofs:sampled}

\textbf{Proof of Theorem~\ref{thm:proposed:quasi-exact}\ref{item:non-empty}).}
For $k = 0$, $\hat T_k = t_k = 0$ and hence $y_k = 0$.
For $k \in \NN$, if $\hat N_k \le L \Delta^2/4$, then $\mathcal{G}_k \cap [1, \gammamax] = \{ 1 \}$; hence $\gamma_k = 1$ and $\hat T_k = \Delta$.
Otherwise, $\Delta/\sqrt{4 \hat N_k/L} \le 1$, and thus the difference between consecutive elements in $\mathcal{G}_k$ is at most one, proving non-emptiness of $\mathcal{G}_k \cap [1,\gammamax]$ with $\gammamax \ge 2$.
In this case, $\gamma_k = j \Delta / \sqrt{4 \hat N_k /L}$ for some $j$, and hence $\hat T_k/ \Delta \in \{k,\kmax,j\} \subset \NN$.
\QED

\textbf{Proof of Theorem~\ref{thm:proposed:quasi-exact}\ref{item:quasi-exact}).}
Let $R \ge 0$ and set $N = 0$ in Proposition~\ref{thm:error:sampled:tight} to obtain $M_0^{L,R}(k \Delta) \le L\Delta/2$ for all $k \in \NN$.
Hence, $M_0^{L,R}(\Delta) = L \Delta/2$ by Proposition~\ref{prop:worst-case-ge-noise-free}.
\QED

\textbf{Proof of Theorem~\ref{thm:sampled:errorband}\ref{item:errorband}).}
The upper bound follows from Proposition~\ref{thm:error:sampled:tight} and the fact that $\gammamax + \gammamax^{-1} \le 2 \sqrt{2}$ for all $\gammamax \in [2, 1+\sqrt{2}]$.
    For $N = 0$, equality to this upper bound follows from Proposition~\ref{prop:worst-case-ge-noise-free}.
    The lower bound is a consequence of Proposition~\ref{prop:sampled-lower-bound} and quasi-exactness shown in Theorem~\ref{thm:proposed:quasi-exact}\ref{item:quasi-exact}).

\textbf{Proof of Theorem~\ref{thm:sampled:errorband}\ref{item:sampled:lower}).}
For $\gammamax \in [2,3/\sqrt{2})$ and with \mbox{$N = L \Delta^2/2$}, one has
    \begin{equation*}
        \hat T_k \le 2 \gamma_k \sqrt{\frac{\hat N_k}{L}} \le 2 \gammamax \sqrt{\frac{N}{L}} = \gammamax \sqrt{2} \Delta < 3 \Delta.
    \end{equation*}
    Hence $\hat T_k \in \{\Delta, 2\Delta\}$ and $k\Delta \ge 2\Delta \ge \hat T_k$.
    Lemma~\ref{lem:better-than-conttime} then yields equality to the lower bound from Theorem~\ref{thm:sampled:errorband}\ref{item:errorband}), i.e., $M_N^{L,R}(k\Delta) =  2 \sqrt{2 N L} - \frac{L \Delta}{2}$.
    \QED

\section{Simulation Results}
\label{sec:simulation}

\begin{figure}
    \centering
    \includegraphics{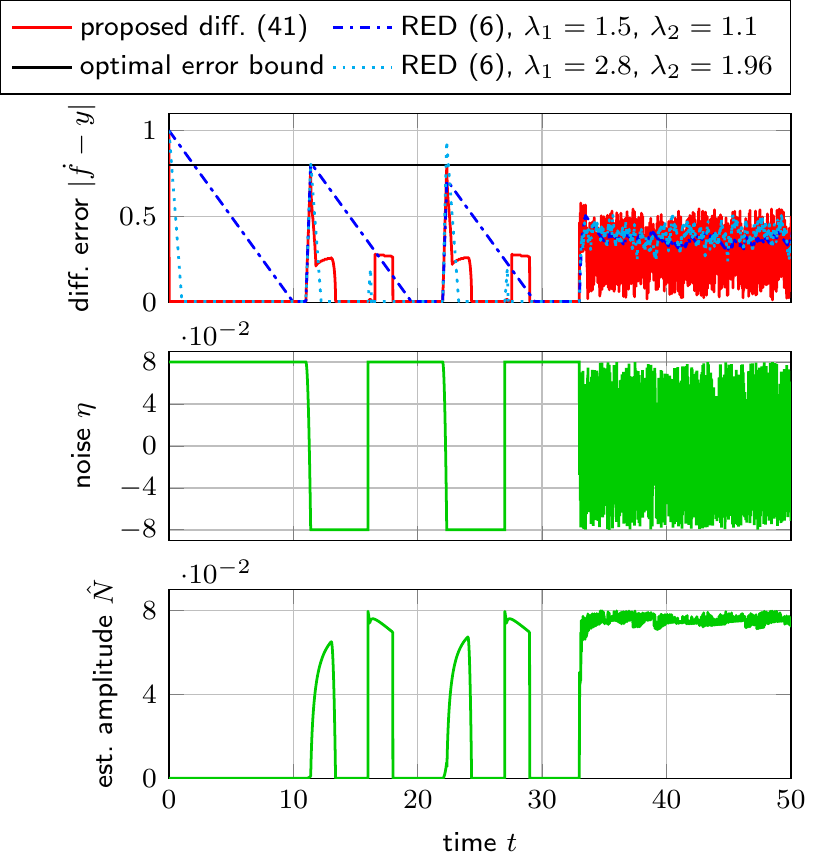}
    \caption{
    Simulation results comparing the RED and the proposed optimal robust exact differentiator with $L = 1$, $N = \changed{0.08}$, and sampling time $\Delta = 0.01$.
    The plots show the differentiation error $|\dot f(t_k) - y_k|$ (top), noise signal $\eta(t_k)$ (center), and noise amplitude estimate $\hat N_k$ (bottom) obtained by the proposed differentiator.
    The optimal error bound is $2 \sqrt{2 N L} = \changed{0.8}$.
    \changed{Maximum errors for $t \ge 10$ of the proposed differentiator, the RED with $\lambda_1 = 1.5$, and the RED with $\lambda_1 = 2.8$ are $0.7939$, $0.8135$, and $0.9374$, respectively.}}
    \label{fig:simulation}
\end{figure}

For illustration purposes, the proposed optimal robust exact differentiator \eqref{eq:diff:sampled} is compared to the sliding-mode based RED \eqref{eq:red}.
A signal $f \in \FL^{R}$ to be differentiated is chosen as $f(t) = L t^2/2 + R t$.
\changed{The differentiators are implemented in discrete time with sampling period $\Delta = 0.01$.
For the RED, the implicit discretization described in \cite{mojallizadeh2021time} is used with differentiator output $y_k = y_{2,k+1}$.
Two different parameter settings are used for the RED: $\lambda_1 = 1.5$, $\lambda_2 = 1.1$ as suggested by \cite{levant1998robust} and $\lambda_1 = 2r$, $\lambda_2 = r^2$ obtained using the tuning procedure used in the toolbox by \cite{andhor_icm21} with a robustness factor of $r = 1.4$.
For the proposed differentiator, the discrete-time realization in \eqref{eq:diff:sampled} is used, where the smallest possible value for $\gamma_k \in \mathcal{G}_k \cap [1,2]$ is selected at every sampling time instant.
Signal and noise parameters are chosen as $L = R = 1$ and $N = 0.08$.
Computational complexity is limited by selecting $\kmax = 200$, corresponding to a continuous-time window length $\Tmax = \kmax \Delta = 2$ and guaranteeing optimal performance for noise amplitudes up to $\Nmax = L \Delta^2 (\kmax - 1)^2/ 2 \approx 1.98$.

Fig.~\ref{fig:simulation} shows the simulation results.
The noise signal is shown in the center portion.
Motivated by Fig.~\ref{fig:exact_worst_case} and practical considerations, it consists of constant segments, two parabola arcs of the principal form $N-(1+\lambda_2)L t^2/2$, with $\lambda_2$ taken from the two RED parameter sets, two step jumps from $-N$ to $N$, and white noise obtained by sampling a uniformly distributed random number from the interval $[-N,N]$.
The top and bottom portions of Fig.~\ref{fig:simulation} depict the differentiation error $|\dot f(t_k) - y_k|$ and the noise amplitude estimate $\hat N_k$.}

\changed{Initially, the noise is constant and the differentiators hence behave as when differentiating a noise-free signal.
One can see that all differentiators make an error equal to $R$ initially; the proposed differentiator then immediately attains quasi-exactness with error bounded by $L \Delta/2$ after a single sampling step, while the 
RED exhibits a finite convergence time, which decreases with increasing $\lambda_2$.

The parabola arcs in the noise lead to peaks in the differentiation error.
With the proposed differentiator, they stay below the optimal worst-case error $2 \sqrt{2 N L}$.
Each RED, in presence of the noise parabola constructed using its respective value of $\lambda_2$, makes an error that exceeds the optimal error bound, with larger values of $\lambda_2$ leading to larger errors.
Tuning of the RED thus requires a tradeoff between worst-case convergence speed and worst-case error, while the proposed optimal robust exact differentiator achieves instant convergence with least possible worst-case error bound.

In the presence of white noise, finally, the frequent step-wise changes in the noise allow for a very accurate estimation of the noise amplitude as predicted by Proposition~\ref{prop:Nhat:properties}.
However, the proposed differentiator also exhibits the largest variation in the error, due to its direct feed-through of the noisy input and the resulting absence of any noise filtering.
The RED, on the other hand, filters high-frequency components of the noise, leading to a smoother error signal overall.
Nevertheless, all differentiators lead to an error of similar magnitude in this case.}

\section{Conclusion}
\label{sec:conclusion}

A first-order differentiator that is robust and exact over a wide class of signals and that achieves optimal differentiation accuracy is proposed for the first time. It is based on the structure of a linear differentiator with a parameter that adapts based on a suitable estimate of the noise amplitude.
It is shown that, in the presence of noise, the proposed differentiator achieves the lowest possible worst-case error among all exact differentiators, and that it converges instantaneously to the true derivative in the absence of noise, hence outperforming all fixed-time convergent differentiators in terms of convergence speed.

For the sampled-data case, a discrete-time implementation of the differentiator based on sampled measurements is provided.
This sample-based differentiator is shown to retain the properties of the optimal continous-time differentiator in their closest possible forms.
In the absence of noise, the proposed sample-based differentiator has the property that it has the least possible worst-case error among all sample-based differentiators; this property is hence called quasi-exactness.
It moreover attains this quasi-exactness after the least possible convergence time of a single sampling step.
In the presence of noise, the worst-case error is shown to converge to a band around its continuous-time optimal value, whose width is as small as possible and is a linear function of the sampling time.

\section*{Acknowledgement}

The authors would like to thank Rodrigo~Aldana-L\'opez, University of Zaragoza, Spain and David~G\'omez-Guti\'errez, Tecnologico de Monterrey, Mexico for several valuable discussions.

\bibliographystyle{abbrvnat}        \bibliography{literature}           \end{document}